\documentclass[aps,reprint,twocolumn,preprintnumbers,floatfix,nofootinbib, amsmath,amssymb]{revtex4-2}

\usepackage{graphicx}
\usepackage{dcolumn}
\usepackage{bm}

\usepackage[utf8]{inputenc} 
\usepackage{color} 
\usepackage[dvipsnames]{xcolor} 
\usepackage{relsize}
\usepackage{graphics}
\usepackage{epstopdf}
\usepackage{hyperref}
\usepackage{mathrsfs}
\usepackage{ragged2e}
\usepackage{amssymb}
\usepackage{placeins}
\usepackage[bottom]{footmisc}

\usepackage[normalem]{ ulem }
\usepackage{amsthm}
\usepackage{amsmath}
\usepackage{hyperref}
\usepackage{cancel}
\usepackage{comment}

\usepackage{nicematrix}

\usepackage{graphics}	 

\begin{document}

\title{Generalising Holographic Superconductors}

\author{Andrea Donini$^a$}%
\author{V\'ictor Enguita-Vileta$^{a,b}$}
\author{Fabian Esser$^a$}
\author{Veronica Sanz$^{a,c}$}

\affiliation{$^a$ Instituto de F\'isica Corpuscular (IFIC), Universidad de Valencia-CSIC, E-46980 Valencia, Spain}
\affiliation{$^b$ Instituto de Física Teórica (IFT), Universidad Autónoma de Madrid-CSIC, E-28049 Madrid, Spain}
\affiliation{$^c$ Department of Physics and Astronomy, University of Sussex, Brighton BN1 9QH, UK}

\date{\today}

\begin{abstract}
In this paper we propose a generalised holographic framework to describe superconductors. We first unify the description of s-, p- and d-wave superconductors in a way that can be easily
promoted to higher spin. Using a semi-analytical procedure to compute  the superconductor properties, we are able to further generalise the geometric description of the hologram beyond the AdS-Schwarzschild Black Hole paradigm, and propose a set of higher-dimensional metrics which exhibit the same universal behaviour. We then apply this generalised description to study the properties of the condensate 
and the scaling of the critical temperature with the parameters
of the higher-dimensional theory, which allows us to reproduce 
existing results in the literature and extend them to 
include a possible description of the newly observed f-wave superconducting systems.
\end{abstract}

\maketitle

\section{Introduction}

In holographic approaches to Superconductivity, the critical behaviour is encapsulated by the dynamics of a dual theory, where interacting fields propagate in the bulk of a fictitious compactified extra-dimension, and whose geometry features dictate the evolution of the system with temperature and its reaction to external electromagnetic fields. Different localisations into the extra-dimensions correspond then to snapshots of the target theory at different energies. This picture is based on the  AdS/CFT correspondence~\cite{Maldacena:1997re,Gubser:1998bc,Witten:1998qj}, the idea that certain strongly coupled theories can be described by an extra-dimensional dual theory which contains weak gravitational interactions.\\
 Based on this framework, models for s-, p- and d-wave superconductivity have been built in the literature~\cite{Hartnoll:2008vx, Hartnoll:2008kx, Gubser:2008wv, Cai:2013aca, Cai:2013pda, Chen:2010mk, Benini:2010pr}, which  provide a computational framework to describe superconducting properties of certain materials.\\
Despite describing a universal phenomenon, criticality, all these models are based on different set-ups and link to different target theories. \\
The aim of this article is to provide a systematic approach to holographic superconductors, which would reveal universal attributes, as well as to put to test some of the features that are commonly introduced in the models. Moreover, this approach could provide insights into models for superconductivity that might be generalised by higher-spin configurations. We will indeed propose  a formulation of Holographic Superconductivity valid for f-wave, g-wave or higher state superconductor. We will further generalise this framework to introduce  extra-dimensional duals which go beyond the AdS-Schwarzschild Black Hole paradigm. These generalisations will be possible in a semi-analytical framework based on matching asymptotic solutions of the differential equations ruling the system.\\
For this matching approach to work, backreactions of the matter fields on the metric are neglected. As it is discussed in Ref.\ \cite{Hartnoll:2008kx}, this limit does not qualitatively change the behaviour of the condensate at $T_c$.

\section{Building a holographic superconductor}
\subsection{The generalised action}
\label{sec:buildingHS:generalised_action}

Holographic superconductivity can be realised in an extra-dimensional setup by enforcing a  field to condense about a Black Hole (BH) horizon in an AdS background.\\
\begin{figure}[h!]
    \centering
    \includegraphics[width=0.45\textwidth]{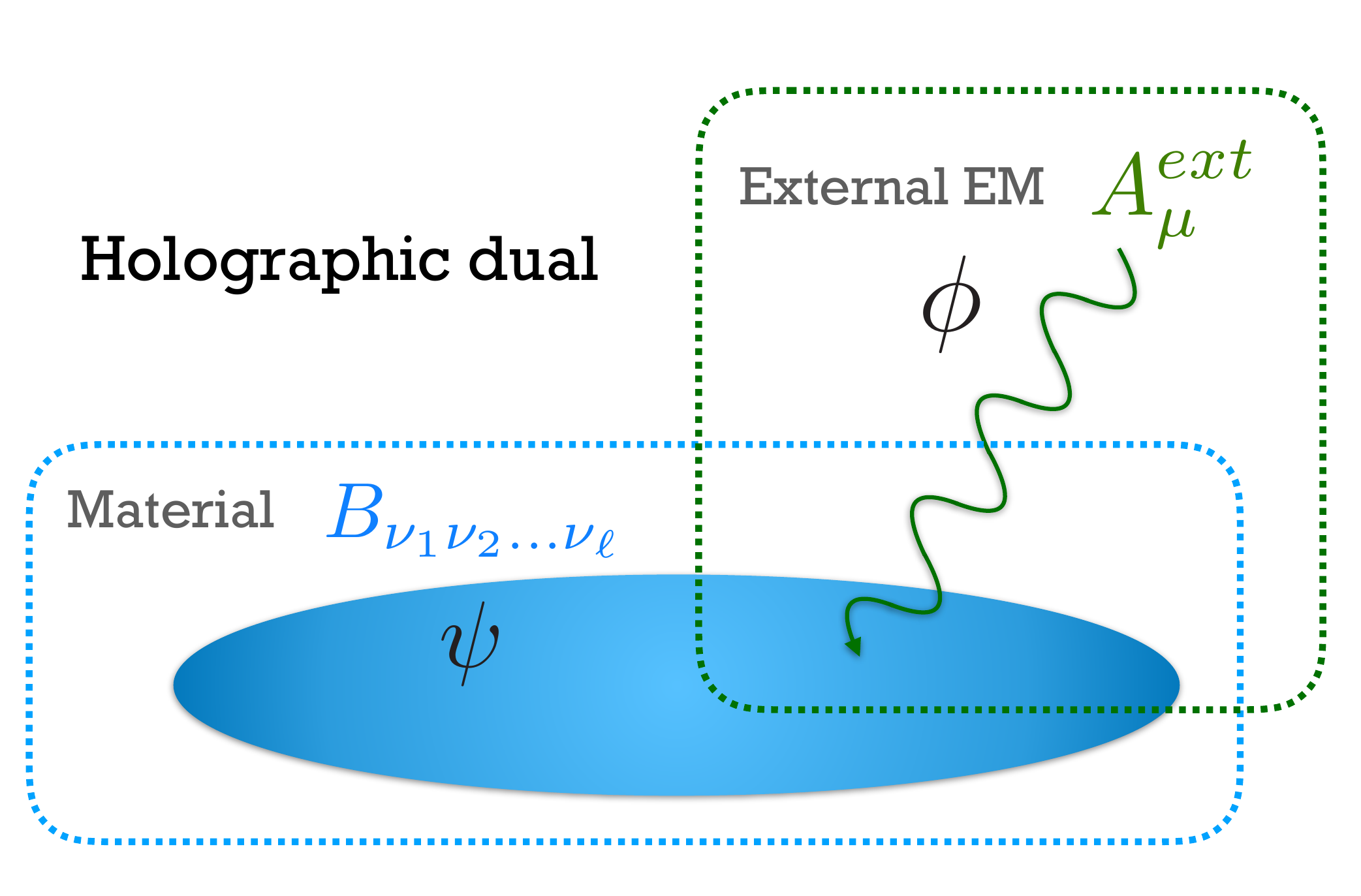}
    \caption{Schematic view of holographic duals to superconductivity: a material represented by a field of spin $\ell$ whose EM properties are probed by an external field living in a higher-dimensional spacetime. The dynamics of these two fields is captured by $\psi$ and $\phi$, respectively.}
    \label{fig:sketch}
\end{figure}
To emulate the electromagnetic properties of the material, the matter field ---in the following the rank $\ell$ tensor  $\boldsymbol{B}_{\nu_1, \dots \nu_\ell}$--- is endowed with a charge $q$ under some $U(1)$ gauge symmetry associated with electromagnetism. Then, both $\boldsymbol{B}_{\nu_1, \dots, \nu_\ell}$ and the $U(1)$ gauge boson $A_\mu$ are minimally coupled to gravity, see Fig.\ \ref{fig:sketch} for a schematic depiction. Further, non-minimal couplings to the metric might as well be included for the purpose of stabilising the theory, in a way that will be shortly discussed. \\
Superconducting theories have been formulated in both 2 and 3 space dimensions, in the former case the materials are layered.  The generalised action that is introduced in this section allows to describe $d$-dimensional superconductors for arbitrary $d$. However, when singling out particular ansatze for the fields, the calculations are performed for layered 2-dimensional superconductors for convenience, where $d$ is then $d=2$+1=3.\\
On describing superconductors in $d$ spacetime dimensions, the  $d$+1-dimensional dual gravitational theory will in general look different depending on the particular realisation. 
But even so, there are a number of common features in every model which this paper aims to point out. The central claim of this work is that the essential dynamical aspects giving rise to a critical behaviour in holographic superconductor theories are captured by a generic action of the form
\begin{equation}
    S= \frac{1}{2 \kappa^2} \int d^{d+1}x \sqrt{-g} \left( \mathcal{R} + \frac{6}{L^2} +\mathcal{L}_m + \mathcal{L}_a\right),
\label{eq:action_general}
\end{equation}
where $\kappa^2 = 8 \pi G_{d}$ is the $d+1$ gravitational strength, $L$ is a length scale that coincides with the AdS radius in a pure AdS metric and the metric signature is $(-,+,+,\ldots)$.
The minimal contributions to the Lagrangian read
\begin{equation}
    \mathcal{L}_m = -\left[\frac{1}
    {4} F_{\mu\nu}F^{\mu\nu} + |D_\mu \boldsymbol{B}_{\nu_1, \dots, \nu_\ell}|^2 + m^2 |\boldsymbol{B}|^2 \right],
\label{eq:general_matter_Lagrangian}    
\end{equation}
where $\mu=0$, 1, 2, $\ldots$ , $d$ is a Lorentz index,  ${D_\mu = \nabla_\mu - iqA_\mu}$, $F_{\mu\nu} = \nabla_\mu A_\nu - \nabla_\nu A_\mu$ and $\nabla_\mu$ is the covariant derivative in the curved spacetime defined by the metric $g_{\mu\nu}$. Meanwhile, the extra piece $\mathcal{L}_a$ is introduced as containing all the extra non-minimal coupling terms between the fields and gravity that might be present in each particular realisation. As it has been discussed \cite{Chen:2010mk, Benini:2010pr}, though not at all essential for the formation of the condensate, the presence of such terms is in general important in safeguarding the stability of the higher-dimensional theory after the introduction of couplings between the matter field, gravity and electromagnetism. More specifically, said stability would require the insertion of an extra set of properly weighted $d+1$ dimensional terms, a subject set beyond the scope of this paper; see Refs.\,\cite{Benini:2010pr, Buchbinder:1999ar} for a deeper insight into the issue for the case $\ell = 2$. \\
A similar approach to Eq.\,(\ref{eq:general_matter_Lagrangian}) has been pursued in Ref.\ \cite{BitaghsirFadafan:2018iqr} to describe Colour Superconductivity, i.e.\ superconductors in the colour pairing of quarks in dense matter QCD.\\
Upon singling out a specific model, the geometrical properties of the superconductivity carriers in the dual Quantum Field Theory (QFT) are determined by the rank of the tensor field $\boldsymbol{B}_{\nu_1, \dots, \nu_\ell}$. In particular, this incorporates the s-, p- and d-wave models present in the literature
\begin{equation*}
    \boldsymbol{B}_{\nu_1, \dots, \nu_\ell}\equiv \begin{cases}
    B &\ell = 0 \, \longrightarrow \, \text{s-wave}\, , \\
    B_\mu &\ell = 1 \, \longrightarrow \, \text{p-wave} \, ,\\
    B_{\mu\nu} &\ell = 2 \, \longrightarrow \, \text{d-wave} \, ,
    \end{cases}
\end{equation*}
together with mixtures of these configurations. Meanwhile, descriptions of higher geometrical complexities (e.g. f-wave with $\ell = 3$) are still lacking. However, also for higher spins $\ell$ the corresponding fields can be described by a rank-$\ell$ tensor with appropriate constraints as we will discuss in Sec~\ref{sec:HigherSpinFields}.\\
Though Eq.\,(\ref{eq:general_matter_Lagrangian}) provides a convenient framework for the study of holographic superconductors in all generality, there are two unsettling aspects of such an action that should be pointed out. Their discussion will be postponed to subsequent sections, but a quick overview of these issues should be of use at this stage already:
\begin{enumerate}
    \item The kinetic part of the minimal Lagrangian in Eq.\,(\ref{eq:general_matter_Lagrangian}) contains not only the physical degrees of freedom but additional unphysical ones that should be removed from the theory for consistency. This is to say, the Lagrangian does not describe by itself the correct number of propagating degrees of freedom.
    \item In curved spacetimes, Eq.\,(\ref{eq:general_matter_Lagrangian}) deviates for $\ell \ge 1$ from known actions for the regarding higher-spin fields because it does not account for curvature corrections.
    This issue will be explained thoroughly in Sec.\,\ref{sec:HigherSpinFields}, where corrected models are presented.\\
    \end{enumerate}
As a result, the theory defined just by the Lagrangian in   Eq.~(\ref{eq:general_matter_Lagrangian}) ought to be distinguished from a full higher-dimensional theory that does not suffer from the mentioned problems,
for which both $\mathcal{L}_m$ and $\mathcal{L}_a$ terms 
in Eq.~(\ref{eq:action_general}) are considered. 
The latter kind of theories will thus from now on be referred to as the \emph{``physical''} ones, whereas the theory that is defined by Eq.\,(\ref{eq:general_matter_Lagrangian}) and bears the mentioned inconsistencies  will receive the label of \emph{``naïve''}. Note, though, that physicality here refers to the higher-dimensional theory we build as a hologram of the {\it real} superconductor system, the target theory. The lack of some features in this higher-dimensional space do not imply an inherent flaw in the target theory, the actual system we want to describe. \\
Nevertheless, two questions immediately arise from this distinction:
\begin{enumerate}
\item Of what use is the \emph{na\"{i}ve} theory, then, if it does not comply with its consistency conditions? 
\item Is there a feasible way to restore physicality? 
\end{enumerate}
In Sec.\,{\ref{sec:HigherSpinFields}} the 
equations of motion (EOM's) for $\ell =0, 1$ and 2 for the \emph{na\"{i}ve} theory will be given explicitly and the inconsistencies in the gravity theory are discussed. Then, it will be shown that the EOM's for some known \emph{physical} theories eventually contain no additional qualitatively relevant contributions when compared to those 
obtained by using the \emph{na\"{i}ve} action.
  
Therefore, an answer can be advanced for the first of these questions: it is expected that the \emph{naïve} model be reliable at the phenomenological level even if there are inconsistencies from the point of view of the gravity theory.
Thus, a study of the formation of the condensate by means of the \emph{naïve} EOM's is justified, and will be carried out in Sec.\,\ref{sec:condensate_analytic_calculations}.\\
The second question is of a rather more technical nature, but as it will be explained in Sec.\,\ref{sec:constraint_equations}, the non-physicalities in Eq.\,(\ref{eq:general_matter_Lagrangian}) can be erased as long as the EOM's that are produced by said action are accompanied by appropriate constraints over the fields. 
This is equivalent to saying that a suitable expression of $\mathcal{L}_a$ is given.
However, while these constraints are easy to grasp in some limits ---e.g. by anti-symmetrising the kinetic term for $\ell =1$ and fixing $B_{\mu\nu}$ to be Lorentz-symmetric, transversal and traceless for $\ell =2$  in the free limit--- the general case for higher-spin and interacting fields is more involved and not iteratively generalisable. Already the formulation of the constraint equations for $\ell = 3$ is well beyond the scope of this work.
The correct formulation of such constraints is in fact an open problem in the field \cite{Buchbinder:1999ar}, for which solutions may only exist in a limited number of background metrics. Nevertheless, the issue is worthy of consideration and will be tackled to some extent in section~\ref{sec:HigherSpinFields}.

\subsection{Ansatze for the fields}
\label{sec:ansatze_fields}

Once the theory has been established, particular ansatze must be implemented for the fields. These ansatze are required to be in accordance with the constraints imposed over the fields that were just discussed in the paragraph above. Typically, the gauge field only depends on the holographic coordinate $r$ and is parametrised as
\begin{equation}
    A_{\mu} dx^{\mu} = \phi(r)dt.
    \label{eq:gaugefixing}
\end{equation}

Regarding the matter field, the different tensor components are fixed so that a single AdS-radial degree of freedom $\Psi(r)$ remains while the required geometrical properties are reproduced. The particular ansatze adopted in this article are shown in Table \ref{tab:MatterTensorRanks}. Note that in the s-wave ($\ell=0$) case spatial isotropy is preserved in the QFT dual, whilst by picking the $B_x$ component in the p-wave ($\ell = 1$) case generates a preferred direction. Finally, this particular ansatz for the d-wave superconductor ($\ell = 2$)\,\cite{Chen:2010mk} generates a condensate in the x-y plane with translation invariance, that breaks the rotational symmetry down to $Z_2$, with the condensate flipping its sign under $\pi/2$ rotations on that plane.

\begin{table}[h!]
    \centering
    \begin{tabular}{|cc|c|c|c|}
    \hline
         Notation & & $\ell$ & $\eta$ & Radial ansatz  \\
    \hline
        $\boldsymbol{B}_
        {\mu_1, \dots \mu_\ell}\equiv $ & $B$ &  0 & 1 & $B = \Psi(r)$ \\
         & $B_\mu$ & 1 & 1 & $B_x = \Psi(r)$\\
         & $B_{\mu\nu}$ & 2 & 2 & $B_{xx} = -B_{yy} = \Psi(r)$ \\ 
         & $B_{\mu\nu\rho}$ & 3 & 2 & $B_{xxx} = - B_{yyy} = \Psi(r)$ \\
         \hline
    \end{tabular}
    \caption{Convenient ansatze for the matter fields in the s- \cite{Hartnoll:2008vx, Hartnoll:2008kx},  p- (adapted from \cite{Gubser:2008wv}), d-wave \cite{Chen:2010mk, Benini:2010pr} and hypothetical f-wave holographic superconductors ($\eta$
    being the multiplicity of $\Psi$ in each case). 
    For each ansatz, every component that is not mentioned vanishes.}
    \label{tab:MatterTensorRanks}
\end{table}

\subsection{The generalised warped metric}
\label{sec:generalised_metric}

It has already been stated that the superconductivity phase transition can be triggered by an AdS BH background metric in the 4-dimensional gravitational theory. Here, this statement will be briefly clarified and, at last, generalisations to the AdS-Schwarzschild metric will be introduced. 

The background metric under consideration is time-independent and fully symmetric in the QFT spatial dimensions. Therefore, it only depends on the holographic coordinate $r$, and can be parametrised as
\begin{equation}
    ds^2 = -f(r) dt^2 + \frac{dr^2}{f(r)} + r^2(dx^2+dy^2),
\label{eq:ansatz_metric}
\end{equation}
In the following, the probe limit $q \to \infty$ is assumed, meaning that the matter fields do not backreact on the metric. In this case, the matching approach, that relies on asymptotic expansions of the fields, provides analytical solutions for the condensates that will be computed further on. \\
Although the probe limit is known to be not numerically accurate for $T/T_C \to 0$, the scaling of the condensate approaching the critical temperature is not affected. This was shown by comparing the probe limit to the numerically obtained fully backreacting solutions e.g. in \cite{Hartnoll:2008kx} for s-wave superconductors. The same applies to the condensate scaling and the dependence of the critical temperature on the charge density for p-wave geometries, as shown in \cite{Arias:2012py, Ghorai:2015wft, Mohammadi:2018ouy, Ammon:2009xh}. \\
Thus, the probe limit is expected to deliver a valid description of the superconductor's properties close to the critical temperature. The main effects due to backreactions are expected to change the value of the condensate at $T/T_C \to 0$, albeit with a lesser impact than the uncertainties arising from the matching point dependence of the semi-analytical approach which is discussed in App.\ref{sec:depzm}. \\
In particular, this means that the metric components are fixed and
$f(r)$ asymptotically becomes the metric of Anti de-Sitter space, i.e. $f \longrightarrow f_{AdS} = r^2$ as $r\longrightarrow \infty$. 

The holographic correspondence identifies QFT's at finite temperature as dual to an AdS space with an event horizon with radius $r_H$, whose introduction renders a physical cut-off in the AdS coordinate. Then, the associated Hawking temperature $T = 3r_H/4\pi$ of the horizon is interpreted as the temperature of the dual QFT. Customarily, for an AdS-Schwarzschild BH located at $r = 0$
the event horizon is set at $r=r_H$, yielding the metric with
\begin{equation}
     f_{BH} (r) = \frac{r^2}{L^2} \left(1- \frac{r_H^3}{r^3}\right),
\end{equation}
which constitutes an exact solution of the gravity EOM inferred from Eq.\,(\ref{eq:action_general}) when the presence of the remaining matter fields is neglected. 

The holographic coordinate is conventionally redefined as $z$\,$=$\,$r_H/r$ which ranges in the interval $z\in (0,1]$ to facilitate the calculations. This means that one usually works with
\begin{equation}
     f_{BH} (z) = \frac{r_H^2}{L^2z^2} \left(1-z^3 \right).
\end{equation}
Notice also that the AdS-asymptotic region is located near $z = 0$ while the horizon lies at $z= 1$.

While the former is a well-motivate choice for the metric, it is not the only one that could be made. For this reason, in this paper other possibilities will be explored with the motivation of continuing the study of holographic superconductors in full generality. This generalisation will be implemented via the ansatz 
\begin{equation}
    f_h(z) = \frac{r_H^2}{L^2} \, h(z) \, (1-z),
    \label{eq:generalized_f_metric}
\end{equation}
that retains explicitly the essential feature of the AdS-BH metric $f_{BH}$, i.e.\ the first-order zero at the horizon with 
\begin{equation}
     h (z)_{AdS} =  \frac{\left(1+z+z^2 \right)}{z^2}.
\end{equation}
Notice that recovering flat AdS metric in the asymptotic region constrains the function $h(z)$ to fulfil $h(z)\rightarrow 1/z^2$ as $z\rightarrow 0$. 

As it will be discussed in Appendix \ref{sec:appendix_generalised_metric}, different choices for the power of $(1-z)$ have been studied but all the cases under consideration failed to reproduce the characteristic temperature behaviour of the condensate. 

Therefore, the compact form of the metric in Eq.~(\ref{eq:generalized_f_metric}) captures the essential features needed to produce a possible holographic model of critical phenomena at finite temperature.

Besides the canonical example of the AdS-Schwarzschild metric, there are various instances of extra-dimensional theories which would be good candidates for a holographic superconductor, while still neglecting backreaction. In particular, the correct pole structure in Eq.~(\ref{eq:generalized_f_metric}) is found on the Hirn-Sanz metrics~\cite{Hirn:2005vk,Hirn:2006wg}, which were employed to develop duals for QCD and for new physics at the Electroweak scale, the AdS-dilaton metric~\cite{Karch:2006pv}, a  specific type of metric designed to reproduce the Regge trajectories, and the Sakai-Sugimoto metric~\cite{Sakai:2004cn,Evans:2008tv},  inspired by Witten's string theory model of D8 branes~\cite{Witten:1998zw}. Specific forms of the function $h(z)$ for these proposals  are summarised below:
\begin{eqnarray}
    h(z) &=& \frac{p_2 (z)}{z^2} \textrm{    (AdS-Schwarzschild~\cite{Witten:1998qj})} \nonumber \\
        &=& \frac{p_2 (z)}{z^2} \, e^{c_d (z/r_H)^{2 d}} \textrm{  (Hirn-Sanz~\cite{Hirn:2005vk,Hirn:2006wg})}\nonumber \\
    &=& \frac{p_2  (z)}{z^2}\, e^{-z^2}\textrm{ (AdS-dilaton~\cite{Karch:2006pv})} \nonumber \\
    &=& \frac{p_{11} (z)}{z^6} \textrm{  (Sakai-Sugimoto~\cite{Evans:2008tv})}\nonumber \\
\label{eq:generalised_metrics:h_examples}
\end{eqnarray}
where $p_n(z) =\sum_{i=0}^{i=n} z^i $ is a polynomial of $z$ up to $z^n$.
Different choices of $h(z)$ would lead to the same universal behaviour but slightly different details. This will be  discussed further in Sec.~\ref{sec:discussion}.

\subsection{Hypothetical higher-order superconductors}

Apart from its compact form that summarises models for s-, p- and d-wave superconductors, the \emph{na\"{i}ve} model presented in Eq.\,(\ref{eq:general_matter_Lagrangian}) allows to speculate about higher order superconductors, for instance about the f-wave superconductor with $\ell = 3$. Indeed, as it will be shown in Sec.\,\ref{sec:condensate_analytic_calculations}, the condensates for the \emph{na\"{i}ve} models share the temperature behaviour with the physical models for $\ell = 0,1,2$. Therefore, it is expected that the \emph{na\"{i}ve} action can provide a valid approximation of the condensates for higher-spin superconductors.  

To give one example, for possible f-wave ($\ell = 3$) superconductors, the matter field is taken to be a rank-3 tensor $B_{\mu\nu\rho}$ with the action
\begin{equation}
    \mathcal{L}_m = -\left[\frac{1}
    {4} F_{\mu\nu}F^{\mu\nu} + |D_\sigma B_{\mu \nu \rho}|^2 + m^2 |B_{\mu\nu\rho}|^2 \right].
\label{eq:Lagrangian_rank3}
\end{equation}
The dynamical field $\Psi$ is singled out by the ansatz $ B_{xxx} = - B_{yyy} = \Psi$ with all remaining components set to zero, compare also Table \ref{tab:MatterTensorRanks}. \\
Note that the mass term in Eq.\,(\ref{eq:Lagrangian_rank3}) in principle would break gauge invariance. Nevertheless, note that this \textit{na\"{i}ve} action should be regarded as an effective description that considers only the relevant degrees of freedom. Some ideas to render the theory consistent will be addressed in Sec. \ref{sec:HigherSpinFields}.

\subsection{The critical temperature and formation of scalar hair}
\label{sec:criticality}

After generalising both the field content  and the geometry of the hologram, we briefly discuss the process of formation of a condensate. 

In the presence of the background metric and the just introduced matter field $B_{\mu_1 ... \mu_\ell}$ with physical components $\Psi(r)$, critical behaviour is triggered by the interplay between electrostatic forces and gravity. The BH attractiveness can be measured in terms of its surface temperature $T$ which is interpreted as the temperature of the dual field theory, too. 

Once equipped with the notion of a finite temperature in the QFT, a critical temperature $T_c$ is determined by the dynamics of the gravity theory such that for $T< T_c$ electrostatic repulsion overcomes gravitational attraction and a superconducting layer can condense over the horizon. For it to actually occur, space-time must be endowed with a negative curvature metric, as it happens in AdS geometry, while in a flat geometry scenario the superconducting layer would be blown out to infinity.

Then, for $T < T_c$, the free energy is minimised by a non-trivial configuration $\Psi = \Psi_0$. This is to say, the $U(1)$ gauge symmetry is spontaneously broken and $\psi$ develops a non-vanishing vacuum expectation value (VEV), leading to the formation of a condensate of the operator in the dual QFT, $\langle \mathcal{O} \rangle \sim \Psi_0$. This can be interpreted as the Black Hole developing scalar hair below $T_c$. 
Meanwhile, for $T>T_c$ the condensate gets swallowed and the system is out of the superconducting phase, the trivial solution $\Psi=0$ is the only viable choice. 

The existence of a critical temperature below which the condensate emerges is the first hallmark of the second order transition giving rise to superconductivity. 

\section{The condensate} 
\label{sec:condensate_analytic_calculations}
 
 \subsection{The generalised EOM's}
Eventually and under the ansatze in Table.~\ref{tab:MatterTensorRanks}
and Eq.~(\ref{eq:gaugefixing}), the theory contains two dynamical scalar fields, $\Psi(r)$ and $\phi(r)$, besides gravity. Under the assumption of a fixed metric ---i.e. neglecting backreactions of these fields upon the metric  itself--- a set of two EOM's can therefore be produced from the action. 

For the class of models defined by Eq.\,(\ref{eq:general_matter_Lagrangian}), the ansatze in Table\,\ref{tab:MatterTensorRanks} and the metric given by Eq.\,(\ref{eq:ansatz_metric}) with $f(z)$ described by Eq.\,(\ref{eq:generalized_f_metric}), a closed expression for an arbitrary tensor of rank $\ell$ can be written down for the``minimal'' part of the action density, $S_m = \sqrt{-g} \mathcal{L}_m$. It is the simplest in terms of the AdS variable $r$ and reads
\begin{equation}
\begin{split}
    S_m &= \frac{1}{2}z^2 \phi'^2 - \eta \left[\frac{f}{r^{2\ell-2}} \Psi'^2 - \frac{2\ell f}{r^{2\ell-2}} \Psi \Psi'\right. \\
    &\left.+ \left( \frac{\ell (\ell + 1 ) f}{r^{2\ell}} +\frac{m^2}{r^{2\ell-2}}  - \frac{q^2}{r^{2\ell-2}f} \phi^2 \right)\Psi^2 \right],
\end{split}
\label{eq:action_UMF}
\end{equation}
where the factor $\eta$ denotes the multiplicity of the dynamical field $\Psi$ within the tensor field itself, cf.\ Table \ref{tab:MatterTensorRanks}.\\
A convenient reparametrisation, inspired by Ref.\,\cite{Benini:2010pr}, consists in writing
\begin{equation}
    \Psi(z) = \frac{\psi(z)}{z^\ell}.
\label{eq:psi_rescaling}
\end{equation}
This allows to write the EOM's in the compact form below:
\begin{equation}
        \phi''(z) -  \left(\frac{r_H}{z}\right)^{2-2\ell}\frac{2\eta q^2\psi^2(z)}{z^2f(z)} \phi(z) =0,
\label{eq:eom_phi}
\end{equation}
\begin{equation}
    \psi''(z) + \frac{f'(z)}{f(z)}  \psi'(z) + M_\psi^2(z)\psi(z)=0 \, ,
\label{eq:eom_psi}
\end{equation}
where a dynamical mass for the $\psi$ field has been defined as
\begin{equation}
M_\psi^2(z) = \frac{q^2 r_H^2 \phi^2(z)}{z^4f^2(z)}-\frac{ r_H^2m^2}{z^4f(z)} - \frac{\ell}{z^2} \ .
\label{eq:dynamical_mass_psi}
\end{equation}
 It must be highlighted that these equations are identical to those in Refs.\,\cite{Hartnoll:2008kx, Gubser:2008wv, Chen:2010mk, Benini:2010pr}, save for the explicitly $\ell$-dependent term $M^2_\psi(z)$. Thus, the difference between the naïve models and those in the literature can be absorbed into the dynamical mass for $\psi(z)$. \\
 Taking into account the rescaling of $\Psi$ in Eq.\,(\ref{eq:psi_rescaling}) the EOM's for p-wave superconductors also agree with those in \cite{Mohammadi:2019swg}.
 
At this point, the EOM's above are all that is needed to unravel the dynamics of the presented generalised holographic superconductor. Though their exact analytical solution is not known, in the remainder of this section it will be shown that, by using the semi-analytical matching procedure presented in Ref.\,\cite{Gregory:2009fj}, one can capture the correct behaviour of the condensate in every case. The reliability of this method for the analysis of the equations under considerations 
ascribes to the absence of important features between $z=1$ (horizon) and $z=0$ (AdS-asymptotic region), and involves the calculation of asymptotic solutions near both regions, that are then matched together at an intermediate point.

\subsection{Solutions near the horizon}
\label{sec:solution_horizon}

The construction of semi-analytical solutions requires the implementation of boundary conditions near the horizon that enforce regularity of Eqs.\,(\ref{eq:eom_phi}) and\,(\ref{eq:eom_psi}). That being so, the former equation imposes
\begin{equation}
    \phi(1) = 0
    \label{eq:horizon:PhiBoundaryCondition}
\end{equation}
on the gauge field $\phi(z)$, while cancelling all divergences in the latter requires that
\begin{equation}
    \psi'(1) = - \frac{\tilde{m}^2}{h(1)} \psi(1).
\label{eq:horizon:PsiBoundaryCondition}
\end{equation}
For convenience, the mass is expressed in units of the curvature from now on, with the dimensionless parameter $\tilde{m}^2 =m^2 L^2$.

Assuming that both $\psi$ and $\phi$ are regular for $z \to 1$ and
keeping terms up to second order in $(1-z)$, one can expand the fields around the horizon in the following manner:
\begin{equation}
    \kappa_H = \kappa(1) - \kappa'(1) \, (1-z) + \frac{1}{2} \, \kappa''(1) (1-z)^2 + \dots
\label{eq:horizon:AnsatzTaylorExpansion}
\end{equation}
for $\kappa \in \{\phi, \psi\}$. The second derivatives at $z=1$ are recovered by plugging these expansions into the EOM's. Substituting the results back into the expansion yields the asymptotic values of the fields in terms of the two still unset boundary values $\phi'(1)$ and $\psi(1)$,
\begin{equation}
    \begin{split}
        \phi_{H}(z) &= -\left[ 1-z + \mathcal{A}_{\ell}\frac{\psi^2(1)}{r_H^{2\ell}}  (1-z)^2\right]\phi'(1) + \dots
    \end{split}
\label{eq:horizon:PhiTaylorExpansion}
\end{equation}
and
\begin{equation}
    \begin{split}
        \psi_{H}(z) = & \left[ 1- \mathcal{B}_{\ell}(1-z) \right.\\ 
        &+ \left. \left(\mathcal{C}_{\ell,1} + \mathcal{C}_{\ell,2} \frac{\phi'^2(1)}{r_H^2}\right)(1-z)^2 \right]\psi(1) + \dots 
    \end{split}
\label{eq:horizon:PsiTaylorExpansion}
\end{equation}
where the following set of coefficients has been conveniently defined:
\begin{equation}
   \mathcal{A}_{\ell} = \frac{q^2\eta}{h(1)},
\label{eq:horizon:TaylorCoeff_A}
\end{equation}
\begin{equation}
    \mathcal{B}_{\ell} = -\frac{\tilde{m}^2}{h(1)},
    \label{eq:horizon:TaylorCoeff_B}
\end{equation}
\begin{equation}
\begin{split}
    \mathcal{C}_{\ell,1} = \frac{1}{4} &\left[-\ell+ 4 \frac{ \tilde{m}^2}{h(1)} + \frac{\tilde{m}^4}{h^2(1)} + 2h'(1)\frac{\tilde{m}^2}{h^2(1)}   \right],
\end{split}
\label{eq:horizon:TaylorCoeff_C1}
\end{equation}
\begin{equation}
    \mathcal{C}_{\ell,2} = -\frac{q^2}{4h^2(1)}.
\label{eq:horizon:TaylorCoeff_C2}
\end{equation}

\subsection{Solutions in the asymptotic region}
\label{sec:solution_asymptotic}

At the AdS boundary ($z=0$), $\psi(z)$ must vanish and $\phi(z)$ be finite to ensure normalisability of the fields. Therefore, the equations for $\psi(z)$ and $\phi(z)$ asymptotically decouple and an independent expansion for each field can be constructed. In the case of the gauge field, one can easily infer that
\begin{equation}
    \phi_{AdS}(z) =\mu - q_\rho z,
\label{eq:asymptotic:PhiTaylorExpansion}
\end{equation}
 where $q_\rho = \rho/r_H$ and $\mu$ and $\rho$ are identified with the chemical potential and the charge density at the horizon, respectively. Meanwhile, since the metric becomes asymptotically AdS, the matter field $\psi(z)$ behaves as
\begin{equation}
    \psi_{AdS}(z) = \mathcal{K}_{\ell} z^{\Delta_{\ell}}
    \label{eq:asymptotic:PsiTaylorExpansion}
\end{equation}
where $\mathcal{K}_{\ell} \propto \langle\mathcal{O}_\ell\rangle$ defines the condensate of the dual operator corresponding to the matter field, denoted by $\mathcal{O}_\ell$, and $\Delta_\ell$ is the mass scaling dimension of said field. \\
The latter can be readily determined to be
\begin{equation}
    \Delta_\ell = \frac{3}{2}  + \sqrt{ \frac{9}{4} + \ell + \tilde{m}^2}.
\label{eq:Delta_ell}
\end{equation}
Note that the definition of $\Delta_{\ell}$ in this work might differ from the convention in other articles. For convenience, the matter field $\Psi$ has been rescaled in Eq.\,(\ref{eq:psi_rescaling}) to give easier matching expressions. But accordingly, the scaling of $\psi$ and $\Psi$ with $z$ differs by a power of $\ell$. Having this transformation in mind, Eq.\,(\ref{eq:Delta_ell}) also perfectly agrees with Eq.\, (13) in \cite{Mohammadi:2019swg} for $d=4$.

\subsection{The matching procedure}
\label{sec:matching}

At last, the matching is realised by imposing Dirichlet and von Neumann boundary conditions at some intermediate point $0<z_M<1$ over each pair of asymptotic solutions. This gives rise to a set of four algebraic equations,
\begin{equation}
\left \{
    \begin{array}{lll}
        \phi_H \left(z_M\right) &=& \phi_{AdS} \left(z_M\right)\, ,\\
        && \\
         \phi'_H \left(z_M\right) &=&  \phi'_{AdS} \left(z_M\right)\, , \\
    \end{array}
\right .
    \label{eq:matching:PhiEquations}
\end{equation}
\begin{equation}
\left \{
    \begin{array}{lll}
         \psi_H\left(z_M\right)
         &=& \psi_{AdS}\left(z_M\right)\, , \\
         && \\
         \psi'_H\left(z_M\right) &=&
         \psi'_{AdS}\left(z_M\right) \, ,
    \end{array}
\right .
\label{eq:matching:PsiEquations}
\end{equation}
whereby the four remaining free parameters $\mu$, $\mathcal{K}_\ell$, $\psi(1)$ and $\phi'(1)$ can be uniquely determined. \\
As demonstrated in Ref.\,\cite{Gregory:2009fj}, the solutions depend only mildly on the precise choice of $z_M$ provided that it is far from its two limiting values, see Appendix~\ref{sec:depzm} for more details. That being so, the convenient choice $z_M = 1/2$ will be adopted in the following. \\
Defining the critical temperature
\begin{equation}
    T_c = \frac{3\sqrt{\rho}}{4\pi} \sqrt[4]{\frac{(2+ \Delta_\ell) \mathcal{C}_{\ell,2}}{2 \mathcal{B}_\ell \left(\Delta_\ell +1 \right)- 4 \Delta_\ell + \left(2+ \Delta_\ell \right)\mathcal{C}_{\ell,1}}}
\label{eq:matching:CriticalTemperature}
\end{equation}
allows to write the solution for $\mathcal{K}_\ell$ for the generalised model under consideration as
\begin{equation}
\begin{split}
   \langle \mathcal{O} \rangle &= r_H^{\Delta_\ell} \mathcal{K}_\ell\\
   =&  W_\ell T_c T^{\ell+\Delta_\ell-1}\sqrt{1+\frac{T}{T_c}}\sqrt{1-\frac{T}{T_c}} \, ,
 \end{split}
\label{eq:matching:CondensateFuncTemperature}
\end{equation}
where one can write
\begin{equation}
    W_\ell = \frac{2^{\Delta_\ell-1}}{\sqrt{ \mathcal{A}_\ell}} \frac{4- \mathcal{B}_\ell}{\Delta_\ell + 2} \left(\frac{4\pi}{3}\right)^{\ell+\Delta_\ell} \, .
    \label{eq:matching:CondensateCoefficient}
\end{equation}

\subsection{Discussion}~\label{sec:discussion}

Eqs.\,(\ref{eq:matching:CriticalTemperature}), (\ref{eq:matching:CondensateFuncTemperature}) and\,(\ref{eq:matching:CondensateCoefficient}) are the main results of this paper, and can be used to describe phenomenological aspects of holographic superconductors, including the parametric dependence of the condensate and critical temperature.

To begin with, as the equations explicitly show, the scaling of the condensate with temperature is of the form
\begin{equation}
    \langle \mathcal{O} \rangle \sim \sqrt{1-\frac{T}{T_c}}
    \label{eq:matching:scaling_condensate}
\end{equation}
for $T\longrightarrow T_c$, exactly as it is expected from a superconducting system. This is a result which was obtained in Ref.\,\cite{Gregory:2009fj} for s-wave superconductors and in an AdS-BH metric, but in this paper it has been shown that such a dependence emerges also for higher-order superconductors and for more general metrics.

The analytical derivation of this scaling relation with the temperature of the condensate, Eq.\,(\ref{eq:matching:scaling_condensate}), is worth describing in more detail:
this scaling behaviour has been shown to be completely determined by the equations for $\phi$, Eq.\,(\ref{eq:matching:PhiEquations}), and is independent of the equations for $\psi$, Eq.\,(\ref{eq:matching:PsiEquations}). This is to say, to this level, the square root behaviour of the condensate appears solely because of the dynamics of $\phi$ and is seen as a consequence of the equilibrium between gravity and the electromagnetic force (cf.\ Sec.\,\ref{sec:criticality}). Qualitatively, one can understand this behaviour by the distinction: the field $\psi$, that carries the information about the material, determines the scale of critical behaviour $T_c$ and the external EM field $\phi$ carries the information on the dependence of the condensate with this temperature. Schematically, 
\begin{equation*}
    \begin{split}
      \psi-\text{dynamics} &\longrightarrow \text{determine } T_c\\
    \phi-\text{dynamics} &\longrightarrow \langle \mathcal{O} \rangle \propto (1-T/T_c)^{1/2}.
    \end{split}
\end{equation*}

Additionally, Eqs.\,(\ref{eq:matching:CondensateFuncTemperature}) and\,(\ref{eq:matching:CondensateCoefficient})  
retain their explicit dependence on the parameters of the gravity theory, namely 
\begin{itemize}
    \item the mass in units of the curvature radius $L$: $\tilde m^2$ 
    \item the free parameters of the generalised metric: $h(1)$ and $h'(1)$.
\end{itemize}  

\begin{figure*}[ht]
    \centering
    \includegraphics[width=\textwidth]{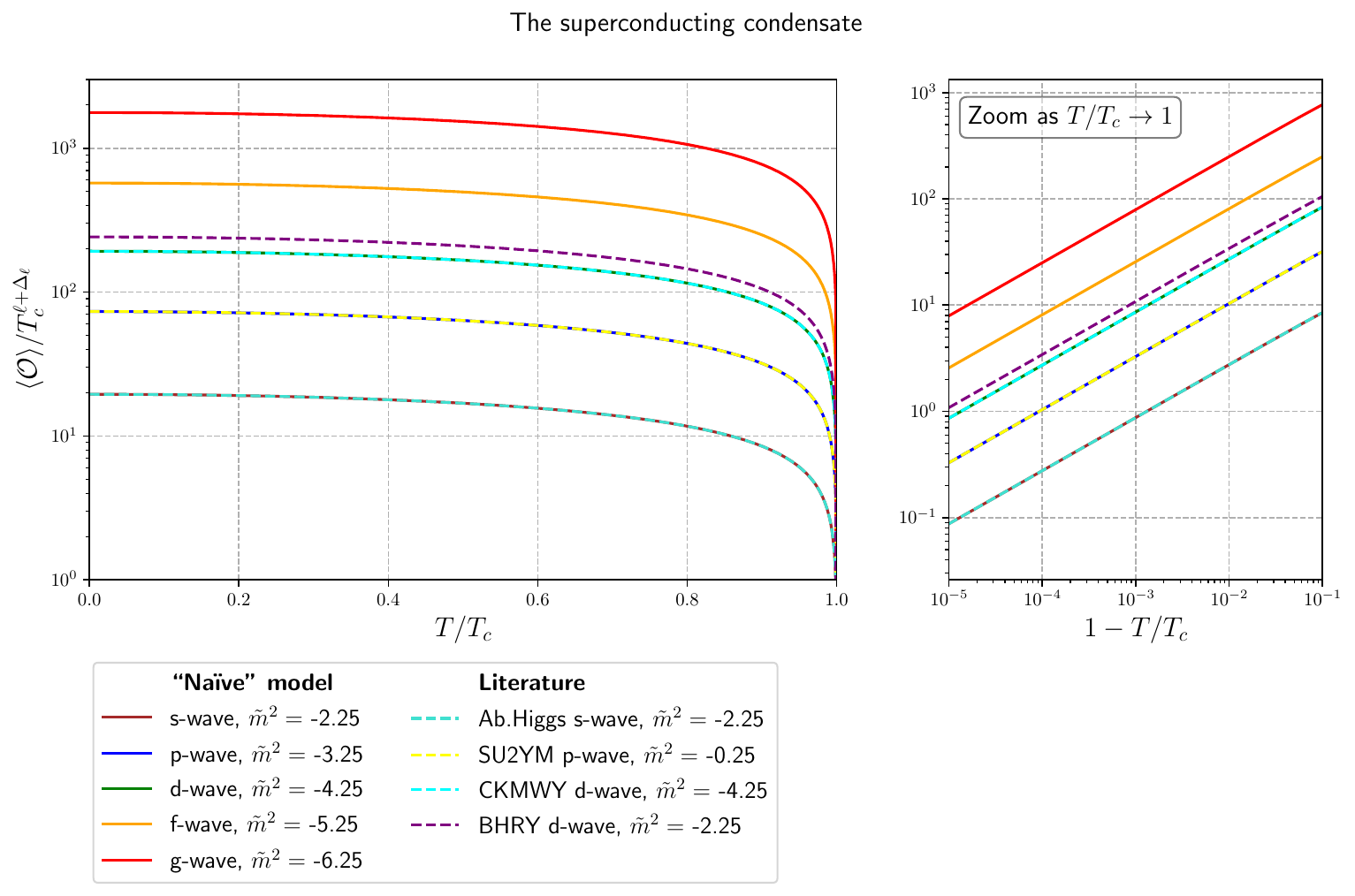}
    \caption{\emph{na\"{i}ve} models for s-, p-, d-, f- and g-wave superconductors and comparison with the Abelian Higgs s-wave, SU(2) Yang-Mills p-wave and CKMWY as well as BHRY d-wave superconductor. All masses are chosen at the respective Breitenlohner-Freedman bound such that the anomalous dimension $\Delta_\ell$ agrees in all cases.}
    \label{fig:condensates_generalised_model}
\end{figure*}
\begin{figure*}[ht]
    \centering
    \includegraphics[width=\textwidth]{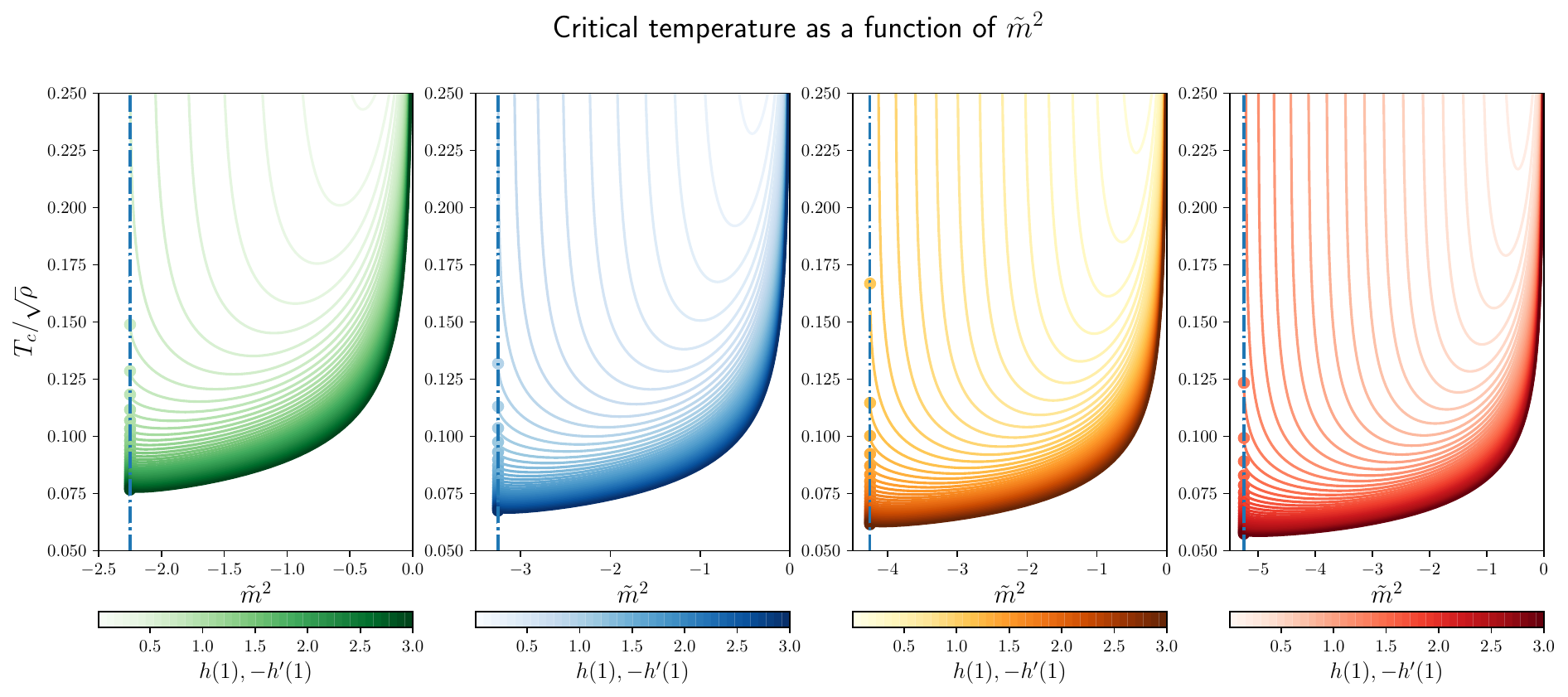}
    \caption{Critical temperature in units of the charge density $\rho$ for s-, p-, d- and f-wave superconductors as a function of $\tilde{m}^2$ and $h(1)$ under the assumption that $h'(1) = -h(1)$. Different curves denote different choices for $h(1)$, the prefactor of the metric at the horizon. Brighter colours refer to small values of $h(1)$ close to zero, darker colours to $h(1)$ close to 3. The dashed blue vertical line symbolises the Breitenlohner-Freedman bound, Eq.\,(\ref{eq:BF_bound}), for each model.}
    \label{fig:t_crit_generalised_models}
\end{figure*}
So, they provide detailed analytical information about the formation of the condensate from the gravity side and allow for a study of this parameter dependence. Notice in fact that a well-defined condensate requires
\begin{equation}
    \{T_c,\langle\mathcal{O}\rangle\}\in \mathbb{R}^+,
\end{equation}
so any point in the parameter space $\{\tilde m^2, h(1), h'(1)\}$ that does not produce this required outcome will not give rise to a superconducting behaviour. 
In particular, notice that rendering $\langle \mathcal{O}\rangle$ and $T_c$ real and positive requires the following conditions to be fulfilled:
\begin{itemize}
\item $\mathcal{A}_\ell \neq 0$ and $\mathcal{C}_{\ell,2} \neq 0$, which are automatically satisfied for non-vanishing $q^2/h(1)$; those conditions simply encode the necessity of a coupling between $\psi$ and the gauge field $\phi$ that is not negligible with respect to the gravitational pull of the event horizon.
\item $\mathcal{B}_\ell <4$ which imposes a relation between $\tilde m^2$ and $h(1)$:
\begin{equation}
    \tilde{m}^2 > -4h(1).
\end{equation}
\end{itemize}
Notice also that the positive definiteness of Eq.\,(\ref{eq:Delta_ell}) prevents approaching the ill-defined limit $\Delta_\ell = -2$~\cite{Breitenlohner:1982bm}. 
Moreover, also from the mentioned equation one can infer that the requirement $\Delta_\ell \in \mathbb{R}$ implies a lower bound for the mass, 
\begin{equation}
    \tilde{m}^2 > - \frac{9}{4} - \ell,
\label{eq:BF_bound}
\end{equation}
which is not but the Breitenlohner-Freedman (BF) bound~\cite{Breitenlohner:1982bm} for this theory. \\
Fig.\,\ref{fig:condensates_generalised_model} shows the dependence of the different condensates with the temperature in units of $T_c$ as described by Eq.\,(\ref{eq:matching:CondensateFuncTemperature}), and puts them in comparison with the condensates obtained from some of the models in literature. \\
Choosing the masses for all models at their respective BF bound, Eq.\,(\ref{eq:BF_bound}), the anomalous dimension is $\Delta_\ell = \frac{3}{2}$ in all cases (cf.\ Eq.\,(\ref{eq:Delta_ell})). Then, for $T$ near $T_c$, the condensates in Eq.\,(\ref{eq:matching:CondensateFuncTemperature}) can be approximated by
\begin{equation}
  \langle \mathcal{O} \rangle  \approx W_\ell T_c^{\ell + \Delta_\ell} \sqrt{1- \frac{T}{T_c}}.
\label{eq:condensate_near_Tc}
\end{equation}
After normalising by $T_c^{\ell + \Delta_\ell}$, the condensates have the same dimension for all values of $\ell$ and differ only by the coefficient $W_\ell$. \\
The right panel focuses on the scaling of either curve near the critical temperature in a double-logarithmic form. From Eq.\,(\ref{eq:condensate_near_Tc}) one reads of the slope in the double-logarithmic form being $1/2$ while the offset is shifted by $\ln W_\ell$.\\
For further comparison between the \emph{na\"{i}ve} and literature models, see Sec.\,\ref{sec:comparison_naive_physical}. Moreover, see Sec.\,\ref{sec:HigherSpinFields} for an outline of the essential features of the latter models together with some of their technical aspects.\\
It remains to look at some of the properties and parameter dependencies of the critical temperature as defined in Eq.\,(\ref{eq:matching:CriticalTemperature}). Fig.\,\ref{fig:t_crit_generalised_models} exhibits the scaling of $T_c$ with $\tilde{m}^2$ and $h(1)$ for s-, p-, d- and f-wave superconductors (from left to right), under the assumption that as in the Schwarzschild BH case $h(1) = - h'(1)$. This is a simplifying assumption that is required in order to reduce the amount of parameters, but the result should not differ much from the general case.\\
For values of $h(1)$ sufficiently far away from 0,
$T_c$ demonstrates a simple scaling with $\tilde m^2$, and is well defined by its only pole at $\tilde m^2 = 0$ all the way down to the BF bound. However, for $h(1)$ approaching zero, it displays a more interesting behaviour. In this regime, the leading contributions to the critical temperature in Eq.\,(\ref{eq:matching:CriticalTemperature}) are given by the $\propto$\,$1/h(1)^2$ terms within $\mathcal{C}_{\ell,1}$ and $\mathcal{C}_{\ell,2}$. Correspondingly, it simplifies to
\begin{equation}
    \lim_{h(1) \to 0} T_c = \sqrt{\frac{1}{\tilde{m}^2 \left( \tilde{m}^2 + 2 h'(1) \right)}}.
\end{equation}
Therefore, the critical temperature visibly diverges both when $\tilde{m}^2\longrightarrow - 2 h'(1)$ and $\tilde{m}^2 \longrightarrow 0$. This behaviour can be spotted in Fig.\,\ref{fig:t_crit_generalised_models}, where the lighter lines corresponding to smaller values of $h(1)$ diverge long before reaching the BF-bound. \\
It is worth noticing the different parametric dependence of $h(1)$ and $h'(1)$ for the typical AdS-BH case and other proposals in Eq.~(\ref{eq:generalised_metrics:h_examples}), see Table~\ref{tab:nonAdS}. Note that $h'(1)$ is typically negative.\\ 
\begin{table}[h!]
    \centering
    \begin{tabular}{c|c|c|c}
         Model & Refs. & $h(1)$ & $h'(1)$\\\hline\hline
         AdS-Schwarzschild & \cite{Witten:1998qj} & $3$ & $-3$ \\ \hline
         Hirn-Sanz & \cite{Hirn:2005vk,Hirn:2006wg} & $3e^{c_d r_H^{-2d}}$ & $e^{c_d r_H^{-2d}}\left(-1+2d c_d r_H^{-2d}\right)$ \\ \hline
         AdS-dilaton & \cite{Karch:2006pv} & $3/e$ & $-9/e$ \\ \hline
         Sakai-Sugimoto & \cite{Evans:2008tv} & $12$ &  $-6$ \\ \hline
    \end{tabular}
    \caption{Parametric dependence of $h(1)$ and $h'(1)$ for the non-AdS cases in Eq.~(\ref{eq:generalised_metrics:h_examples}).}
    \label{tab:nonAdS}
\end{table}

\begin{figure}[h]
    \centering
    \includegraphics[width=0.5\textwidth]{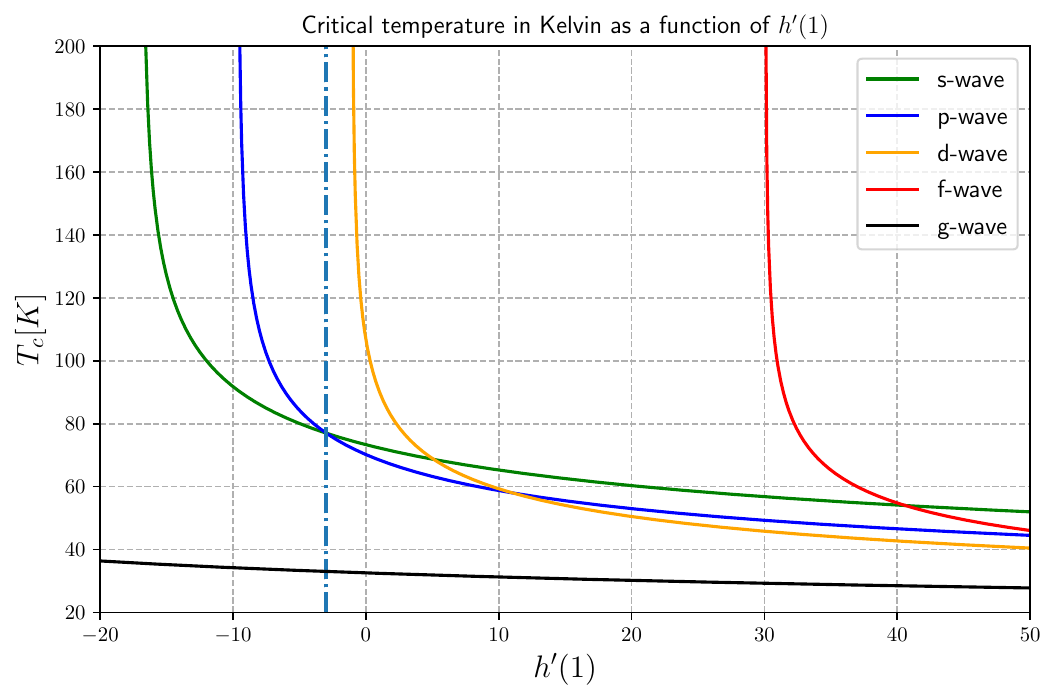}
    \caption{Critical temperature in Kelvin for s-, p-, d-, f- and g-wave superconductors as a function of $h'(1)$ for $h(1) = 3$ and all masses chosen at the respective BF bound, Eq.\,(\ref{eq:BF_bound}). The energy density of the gauge field $\phi$ near the horizon has been set to $\rho$\,$=$\,$7.5$\,$\cdot$\,$10^{-3}$\,$\textrm{eV}^2$ so as to relate the prediction of the model with physical systems. The dashed blue vertical line emphasises the AdS Schwarzschild BH case $h'(1) = -3$.}
    \label{fig:t_crit_generalised_models_Kelvin}
\end{figure}

So far, all temperatures have been presented in units of the charge density $\rho$. However, the distinctive quantity to relate the models to existing or hypothetical physical systems is the critical temperature in Kelvin.
Therefore, the charge density $\rho$ is fixed such that the critical temperature of the s-wave superconductor gives $77$K at the BF bound $\tilde{m}^2 = -2.25$ and for the parameter choices $h(1) = 3$ and $h'(1) = -3$ as in the Schwarzschild BH case. \\
The critical temperature in Kelvin for some choices of $\ell$ obtained from Eq.\,(\ref{eq:matching:CriticalTemperature}) is shown in Fig.\ \ref{fig:t_crit_generalised_models_Kelvin} as a function of $h'(1)$.  
Note that for s-, p-, and g-wave superconductors a finite value of the critical temperature can be defined in the Schwarzschild BH limit $h'(1)=-3$, while it diverges before reaching this limit for d- and  f-wave superconductors. \\
The semi-analytical matching approach pursued in this article requires a mild dependence on the matching point $z_M$. As it will be discussed in Appendix~\ref{sec:depzm}, this is the case for $\tilde{m}^2$ close to the BF bound. 
However, the curves in Fig.\ \ref{fig:t_crit_generalised_models_Kelvin} imply that the AdS Schwarzschild BH background metric does not provide a critical temperature at the BF bound for d- and f-wave superconductors. Accordingly, this fact suggests to take into account generalised background metrics as it has been motivated in Sec.\ \ref{sec:generalised_metric}.\\
Apart from their condensate, holographic superconductors are typically characterised by the optical conductivity. However, the analytical matching approach presented here breaks down for the conductivity, see App.~\ref{sec:appendix_conductivity} for details.

\section{More on higher spin fields} 
\label{sec:HigherSpinFields}

\subsection{Na\"{i}ve versus physical models}
\label{sec:HigherSpin_Naive_vs_Physical}
The \emph{na\"{i}ve} model introduced in Eqs.\,(\ref{eq:action_general}) and (\ref{eq:general_matter_Lagrangian}) and the thereupon derived EOM for $\psi$, Eq.\ (\ref{eq:eom_psi}), bear the two main shortcomings that were already mentioned in Section\,\ref{sec:buildingHS:generalised_action}. First and foremost, the kinetic part of Eq.\,(\ref{eq:general_matter_Lagrangian}) is inconsistent with the axioms of Quantum Field Theory in curved spacetime as it leads to unphysical particle states and it does not describe the correct number of propagating degrees of freedom. Secondly,
Eq.\,(\ref{eq:general_matter_Lagrangian}) does not exactly reproduce the known models for p-and d-wave superconductors and only the s-wave case matches the models in the literature, despite producing a similar superconducting phenomenology. \\
It is the latter problem that can be demonstrated the easiest to tackle, and will be the focus of attention in the following section. \\
It should then be illustrative to enumerate the differences between the generalised, naïve model defined by Eq.\,(\ref{eq:general_matter_Lagrangian}) in the non-interacting limit and the well-established Lagrangians that describe free tensor fields of ranks $\ell = 1$ and $2$ for different spacetimes. 
By doing so, the aims of the writers are two-fold: this discussion shall contribute to explaining where the differences originate from and will hopefully cast some light on some of the issues that could arise for higher spin fields. \\
Rank-$1$ tensors are suitable to describe the matter part of a p-wave holographic superconductor, and the right description of the dynamics of such fields in the free limit for Minkowski spacetimes is known to be given by the Proca Lagrangian\,\cite{Proca:1936ix} 
\begin{equation}
    \mathcal{L}_\text{Proca} =  -\frac{1}{4} \left|\partial_{\mu} B_{\nu} - \partial_{\nu} B_{\mu} \right|^2 - \frac{1}{2} m^2 \left|B_{\nu}\right|^2
\label{eq:Langrangian_pwave_Proca}
\end{equation}
whose mass term would break gauge invariance. Note again that this action should be regarded as an effective action, containing only the relevant degrees of freedom to describe Superconductivity.
However, for p-wave superconductors in flat spacetime and with $q^2 = 0$, the \emph{na\"{i}ve} model reduces to the matter Lagrangian
\begin{equation}
    \mathcal{L}_m \supset   -\frac{1}{2} \left|\partial_{\mu} B_{\nu}\right|^2  - \frac{1}{2} m^2 \left|B_{\nu}\right|^2 
\label{eq:Lagrangian_pwave_UMF}
\end{equation}
that differs from Eq.\,(\ref{eq:Langrangian_pwave_Proca}) by the lack of antisymmetrisation of the kinetic term and leads to the Klein-Gordon like EOM
\begin{equation}
    \left( \Box - m^2\right) B_{\nu} = 0, 
\label{eq:eom_pwave_naive}
\end{equation}
where $\Box$ is the 4-dimensional D'Alembertian operator. \\
The imprint that the mentioned difference between Eq.\,(\ref{eq:Langrangian_pwave_Proca}) and Eq.\,(\ref{eq:Lagrangian_pwave_UMF}) leaves on the dynamics of the theory depends crucially on the considered background spacetime. Note that in the so far considered Minkowski spacetime, the Proca equation inferred from the Lagrangian in Eq.\,(\ref{eq:Langrangian_pwave_Proca}) reads 
\begin{equation}
     \partial_\mu \partial^\mu B_{\nu} - \partial_\nu \partial^\mu B_\mu   - m^2  B_\nu = 0
\label{eq:eom_Proca}
\end{equation}
which simplifies in the Lorentz gauge $\partial_\mu B^{\mu} = 0$ to the EOM obtained from the \emph{na\"{i}ve} action, Eq.\,(\ref{eq:eom_pwave_naive}). This is to say, in Minkowski spacetime, the lack of antisymmetrisation does not change the dynamics. \\
Next, one can see with little effort that turning on the minimal $U(1)$ couplings and promoting the partial derivatives to gauge-covariant ones, ${\partial_{\mu} \rightarrow D_{\mu} = \partial_{\mu} - i q A_{\mu}}$, does not introduce any difference either. This means, minimally coupling the fields in Minkowski spacetime does not introduce any deviation in the dynamics derived from the naïve action with respect to the physical action.

In contrast, covariantising the partial derivatives $\partial_\mu$\,$\rightarrow$\,$\nabla_\mu$ in curved spacetimes, the second term in Eq.\,(\ref{eq:eom_Proca}) no longer vanishes in the Lorentz gauge and introduces additional terms with respect to the naïve EOM. To see this explicitly, one can write the Proca equation for $\Psi$ in presence of the generalised background metric and switch on the minimal couplings to the $U(1)$ gauge field, obtaining
\begin{equation}
    \Psi'' + \left( \frac{f'}{f} + \frac{2}{z} \right) \Psi' + \left( \frac{q^2 \phi^2}{z^4f^2} - \frac{m^2}{z^4 f} \right) \Psi = 0
\end{equation}
that coincides with the SU(2) Yang-Mills model for a p-wave superconductor \cite{Gubser:2008wv}. Meanwhile, the EOM obtained from Eq.\,(\ref{eq:Lagrangian_pwave_UMF}) is\footnote{Notice that trading $\Psi$ for $\psi$ via Eq.\,(\ref{eq:psi_rescaling}), Eq.\,(\ref{eq:eom_psi_pwave}) becomes Eq.\,(\ref{eq:eom_psi}) for $\ell =1$.}
\begin{equation}
    \Psi'' + \left( \frac{f'}{f} + \frac{2}{z} \right) \Psi' + \left( \frac{q^2 \phi^2}{z^4f^2} - \frac{m^2}{z^4 f} + \frac{1}{z} \frac{f'}{f} -\frac{1}{z^2} \right) \Psi = 0
\label{eq:eom_psi_pwave}
\end{equation}
which contains two further contributions to the 0$^\text{th}$ order term in $\Psi$ that originate from derivatives of the metric components and would not be present in flat space. However, it is noteworthy that these two terms enter the equation at the level of the mass term. Consequently, they only affect the definition of the critical temperature but not the shape of the condensate, and do not crucially affect the dynamics of the theory after their presence is accounted for; see Sec. \ref{sec:comparison_naive_physical} for a proof of this fact. As promised, they hence do not alter the phenomenology of the model.\\
In d-wave holographic superconductors, the matter field is described by a rank-2 tensor $B_{\mu\nu}$ whose exact dynamics in the free limit and for a Minkowski background are now given by the Fierz-Pauli Lagrangian\,\cite{Fierz:1939ix}
\begin{equation}
    \begin{split}
        \mathcal{L}_\text{F-P} &= \partial_\rho B\partial^\rho B - \partial_{\rho} B_{\mu\nu}\partial^{\rho} B^{\mu\nu} - 2\partial^{\rho}B_{\rho\mu}\partial^{\mu} B\\ 
        + 2&\partial_{\rho} B_{\nu\mu} \partial^{\nu}B^{\rho \mu} - m^2 \left( B_{\mu\nu} B^{\mu\nu} - B^2 \right)
        \label{eq:Lagrangian:Fierz-Pauli}
    \end{split}
\end{equation}
where the trace $B = \eta^{\mu\nu}B_{\mu\nu}$ is defined using the Minkowski metric $\eta^{\mu\nu}$. The naïve model for vanishing $U(1)$ couplings in Minkowski spacetime yields a rather more succinct expression for $\ell = 2$,
\begin{equation}
    \mathcal{L}_m = - \left|\partial_{\rho} B_{\mu\nu} \right|^2 - m^2 \left|B_{\mu\nu}\right|^2.
    \label{eq:Lagrangian:naive_d-wave}
\end{equation}  
 Now, 
 this Lagrangian turns out to coincide exactly with Eq.\,(\ref{eq:Lagrangian:Fierz-Pauli}) in the traceless and transversal gauge given by
\begin{equation}
{B_\mu}^\mu = 0 \ \ \text{ and } \ \ \partial^\mu B_{\mu\nu} =0.
\end{equation} 
Accordingly, the naïve EOM equals the Fierz-Pauli equation in this gauge, in which they read
\begin{equation}
    (\Box - m^2 ) B_{\mu\nu} = 0.
\end{equation}
Thus, once more, both the physical and the naïve theories turn out to display the same dynamics.

Yet again, difficulties arise when considering curved spacetimes that are even more problematic to address than in the former case. Remarkably and as it is discussed in Ref.\,\cite{Buchbinder:1999ar}, adding $U(1)$ gauge couplings can also introduce instabilities in the $\ell = 2$ theory. However, although this is true on a technical level, it will be shown that the EOMs obtained from the naïve action and the minimally coupled extension of the Fierz-Pauli action still  agree for the chosen ansatz for $B_{\mu\nu}$.

In curved spacetimes, the case is indeed more difficult than that of the p-wave superconductor. In particular, covariantising the Fierz-Pauli equation and introducing $U(1)$ minimal couplings does not suffice to produce the required number of independent constraint equations\,\cite{Benini:2010pr}. Consequently, spurious propagating degrees of freedom are left in the theory that lead to, in principle, disastrous inconsistencies such as the emergence of ghosts or the  loss of hyperbolicity. 

Several approaches exist to cure this kind of difficulties, one of which is built in the BHRY model presented by Ref.\,\cite{Benini:2010pr}. As will be shown in Sec.\,\ref{sec:constraint_equations}, this model is constructed by including in the action all the possible terms of scaling dimension $d+1$ that are quadratic in the spin-2 field, weighted by coefficients that are treated as Lagrange multipliers. By fixing these coefficients one ensures the existence of a correct number of independent constraint equations that retain the physical degrees of freedom only. This process leads to the BHRY Lagrangian that contains a number of non-minimal couplings between the spin-2 field, the gauge field and gravity as well as a more involved kinetic sector:
\begin{equation}
\small
\begin{split}
    \mathcal{L}_{BHRY} = &- \left|D_{\rho} B_{\mu\nu} \right|^2 + 2\left| D_{\mu} B^{\mu\nu}\right|^2 - \left[ D_{\mu} B^{*\mu\nu} D_{\nu} B + h.c. \right] \\
    & + \left| D_{\mu} B \right|^2 - m^2 \left( \left|B_{\mu\nu}\right|^2 - \left|B\right|^2 \right) + 2 R_{\mu\nu\rho\lambda} B^{*\mu\rho}B^{\nu\lambda} \\
    & - R_{\mu\nu} B^{*\mu \lambda} {B^{\nu}}_\lambda
    -\frac{1}{d+1} R \left|B\right|^2 -i q F_{\mu\nu} B^{*\mu \lambda} {B^{\nu}}_\lambda.
\end{split}
\label{eq:Lagrangian:action_BHRY}
\end{equation}
However, in spite of its rather involved form, this Lagrangian simplifies in the traceless and transversal gauge for $B_{\mu\nu}$, in which it reads
\begin{equation}
    \mathcal{L}_{BHRY}^{TT} = - |D_\rho B_{\mu\nu}|^2 - m^2 |B_{\mu\nu}|^2 + 2 R_{\mu\nu\rho\lambda} B^{\mu \rho} B^{\nu \lambda},
\label{eq:action_BHRY_tt_gauge}
\end{equation}
i.e.\ it only differs from the \emph{na\"{i}ve} action by the curvature correction term involving the Riemann tensor $R_{\mu\nu\rho\lambda}$.
The corresponding EOM is
\begin{equation}
    (\Box - m^2) B_{\mu\nu} + 2 R_{\mu \rho \nu \lambda} B^{\rho \lambda} = 0.
\end{equation}
It is more illustrative to compare the equations arising from each model explicitly. Making use of the ansatz for $B_{\mu\nu}$ in Table \ref{tab:MatterTensorRanks}, the \emph{na\"{i}ve} EOM translates into
\begin{equation}
    \Psi'' + \left( \frac{f'}{f} + \frac{4}{z} \right) \Psi' + \left( \frac{q^2 \phi^2}{z^4f^2} - \frac{m^2}{z^4 f} + \frac{2}{z} \frac{f'}{f} \right) \Psi = 0,
\end{equation}
which coincides with that of the CKMWY model in Ref.\,\cite{Chen:2010mk},
while the BHRY equation, corrected by the curvature term $R_{\mu\rho\nu\lambda}B^{\rho \lambda} = \frac{2f}{r^2} \psi$ delivers\footnote{The Riemann tensor for the generalised metric has been calculated with the sub-package \textit{xCoba} of the Mathematica package \textit{xAct}\,\cite{xAct:2002}.}
\begin{equation}
        \Psi'' + \left( \frac{f'}{f} + \frac{4}{z} \right) \Psi' + \left( \frac{q^2 \phi^2}{z^4f^2} - \frac{m^2}{z^4 f} + \frac{2}{z} \frac{f'}{f}  + \frac{2}{z^2}\right) \Psi = 0.
\end{equation}
Hence once more, notwithstanding the formal issues of the spin-2 case, the differences between the models reduce to a series of curvature corrections that enter the EOM's at the level of the dynamical mass. Then, and as it happened in the p-wave case, the corrections only change the definition of the critical temperature and should not affect the model phenomenologically.

Notice also that, in terms of $\psi$, the latter equation takes the simpler form
\begin{equation}
        \psi'' + \frac{f'}{f}  \psi' + \left( \frac{q^2 \phi^2}{z^4f^2} - \frac{m^2}{z^4 f} \right) \psi = 0
\end{equation}
in which the curvature corrections have been absorbed by the redefined field.
One consequence is that after such a redefinition of the field the EOM's of the naïve $\ell = 2$ model are identical to those of the s-wave case, see for instance Ref.\,\cite{Hartnoll:2008kx, Hartnoll:2008vx}.\\

After describing the procedure for  p- and d-wave superconductors, it is time to address the problem for a field of rank $\ell > 2$.  
Even though the correct theory describing the dynamics of such higher spin fields in the general case is still to be formulated, there are some cases that are well known in the literature with which the model presented in this paper can be compared. In fact, the simplest possible situation of a massless higher spin field in Minkowski spacetime is well-understood, the Fronsdal equation\,\cite{Fronsdal:1978rb} providing the right description of the propagating degrees of freedom in that case. It reads
\begin{equation}
    \Box B_{\mu_1 ... \mu_\ell} - \partial_{(\mu_1} \partial^{\sigma} B_{\mu_2 ... \mu_\ell) \sigma} + \frac{1}{2} \partial_{(\mu_1} \partial_{\mu_2} {B_{\mu_3 ... \mu_\ell) \sigma}}^{\sigma} = 0,
\label{eq:EOM:Fronsdal_equation}
\end{equation}
which reduces to $\Box B_{\mu_1 ... \mu_\ell} = 0$ in the traceless and transversal gauge
\begin{equation}
B_{\mu_3 ... \mu_\ell \sigma}^{\sigma}=0 \ \text{ and } \ \partial^{\sigma} B_{\sigma \mu_3 ... \mu_\ell}=0. 
\end{equation}
Trivially, this coincides with the EOM arising from Eq.\,(\ref{eq:general_matter_Lagrangian}) in the massless limit $m^2 = 0$. 

But again, considering curved spacetimes seemingly puts one in dire straits. The main issue in this case is the non-commutativity of the covariant derivatives, which prevents the second term in Eq.\,(\ref{eq:EOM:Fronsdal_equation}) from vanishing and introduces curvature correction terms that correspond to the higher spin generalised version of those that have explicitly been found in the p- and d-wave cases. 

A generalised procedure exists to construct the required set of constraint equations to cure all instabilities in the rank-$\ell$ case. This method will be outlined in Sec.\,\ref{sec:constraint_equations} but it should not be regarded as an universal remedy, as for larger $\ell$ it becomes involved and there is no systematic generalisation of every of its steps. 

Therefore, it would be useful to have
an operative method to calculate these curvature corrections without having to explicitly consider constraint equations. In the AdS background, several worthwhile attempts to describe the dynamics of free higher-spin fields have been made in the literature \cite{Schwartz:2005ir, Karch:2006pv, Afonin:2011ff, Jin:2015aba}. 
One approach displayed in Ref.\, \cite{Kessel:2017mxa} is to regard the curvature correction terms as belonging to a dynamical mass $M^2_\ell$, yielding an equation of the form

\begin{equation}
    \left(\Box - \frac{M_\ell^2}{L^2} \right) B_{\mu_1 ... \mu_\ell} = 0
\label{eq:Fronsdal_effective_mass_AdS}
\end{equation}
in the traceless and transversal gauge, where the mass takes the form
\begin{equation}
    M_\ell^2 = \ell^2 + \ell(D-6) - 2(D-3)
\end{equation}
in $D = d+1$ spacetime dimensions.

For the presented superconducting theories, though, the correct curvature corrections are hard to write down in a general form for higher spins. The difficulty of this enterprise arises from two sources:
\begin{enumerate}
    \item The non-purely AdS metric.
    \item The couplings with the $U(1)$ electromagnetic field.
\end{enumerate}
So as to visualise the first issue, one can consider the EOM for the matter fields in terms of the initial field $\Psi$, i.e. before performing the redefinition of the matter field defined by Eq.\,(\ref{eq:psi_rescaling}). In those terms, the EOM for general $\ell$ takes the form
\begin{equation}
\begin{split}
    \quad \Psi'' + \left(\frac{f'}{f} + \frac{2\ell}{z} \right) \Psi'
     + & M_\Psi'^2(z, \ell) \Psi = 0,
\end{split}
\label{eq:eom_psi_withEandF}
\end{equation}
where the dynamical mass $M_\Psi'^2(z)$ is shifted by a pair of curvature correction terms,
\begin{equation}
M_\Psi'^2(z) = \left( \frac{q^2 r_H^2 \phi^2}{z^4 f^2} - \frac{m^2 r_H^2}{z^4 f} + \frac{ E_\ell f'}{zf} + \frac{F_\ell}{z^2} \right) 
\label{eq:dynamical_mass_E_F}
\end{equation}
and the coefficients $E_\ell$ and $F_\ell$ weighting those terms are
\begin{equation}
    E_\ell = \ell \ \text{ and } \ F_\ell = \ell(\ell-2).
\label{eq:coefficients_E_F}
\end{equation}
Now, in a pure AdS background where $f(z) = r_H^2/z^2$, the $m^2$, $E_\ell$ and $F_\ell$ terms all scale with $z^{-2}$ and can be collected into a single dynamical mass term proportional to
\begin{equation}
m_{dyn}^2 = \tilde{m}^2 + 2E_\ell - F_\ell.
\end{equation} 
Hence the curvature corrections become only a shift to an effective mass and are effectively erased from the EOM's. But unfortunately, this does not hold for arbitrary $f(z)$ where the derivative of the metric $f'(z)$ is not proportional to $f(z)$, spoiling the convenient relation from AdS.
Meanwhile, the coupling to the gauge field $\phi$ also complicates the argumentation and sabotages some ideas.
In fact, one very promising path to present the EOM's in a neat form leads down to exploiting the invariance of an action of the form Eq.\,(\ref{eq:general_matter_Lagrangian}) under gauge transformations $\delta \boldsymbol{B}$ for free fields in AdS-like theories, as shown for instance in Ref.\,\cite{Karch:2006pv}. In that paper it is shown that by electing an appropriate gauge, the fields can be rescaled in such a way that the action only depends on partial derivatives of these fields and the EOM's can directly be read off. But even though this method could in general take care of general metrics by adapting the field rescaling, it is easy to see that it eventually breaks down when including gauge field interactions via gauge-covariant derivatives. 

\subsection{Constructing complete models for higher spin fields}
\label{sec:constraint_equations}

To date no complete, pathology-free formulations of higher spin theories have been found. However, several methods exist in the literature which suggest that doing so is formally possible, at least for a limited number of space-time geometries that include AdS. Therefore, after having pointed out the differences between the naïve and physical models, it will be interesting to provide more detail about how it can be done in the general case by putting together a sufficient set of constraint equations.

The procedure is explained in Ref.\,\cite{Buchbinder:1999ar} and it is summarised here for the sake of completeness. The main idea was already anticipated in the examples in the first part of this Section. While transversality and tracelessness conditions are enough to eliminate unphysical degrees of freedom in free theories, further constraints are required when interactions are switched on in a higher spin field theory. For consistency, as a first requisite one needs that the same number of degrees of freedom as in the flat theory are conserved when introducing interactions. Another crucial point is the preservation of causality i.e., the absence of superluminal motion must also be enforced.

The general algorithm that allows to construct a theory fulfilling those requirements goes at follows: let $N$ be the number of degrees of freedom contained in the \emph{na\"{i}ve} theory, that will be denoted as $\phi^a$ following Ref.\,\cite{Buchbinder:1999ar}. By applying the principle of least action, a set of $N$ equations can be obtained from the theory that are not all independent of each other. In particular, the equations can be rewritten and split in two different categories:
\begin{itemize}
    \item A set of $k\le N$ equations which fix $k$ of the second derivatives $\ddot{\phi}^a$ of the degrees of freedom. 
    \item The remaining $N-k$ equations which do not contain accelerations. These are called \emph{primary constraint equations}.
\end{itemize}
As a consequence, the system of equations that directly arises from the action does not completely characterise the dynamics of the theory and additional constraints must be brought in from elsewhere.
One can recur, at this point, to conservation of the primary constraints. By doing so, a new set of $N-k$ \emph{secondary constraint equations} is found, containing second derivatives which may or not fix the still undetermined degrees of freedom. 

If they do not, one can then iteratively repeat this procedure until all second derivatives are fixed by the equations. Typically, for rank-$\ell$ tensor fields a total of $\ell$ iterations are necessary, which generically require to rewrite the equations in an appropriate form. From this general insight it becomes clear that the actual calculation of the constraints will be complicated for higher spin models and cannot be generalised automatically to arbitrary $\ell$.

For an explicit example, the case of the $\ell =2$ d-wave superconductor will be considered. The problem was in fact solved for this case in Ref.\,\cite{Benini:2010pr} to set up the BHRY superconductor. The model aims to describe a massive spin-2 field in $d+1$ spacetime dimensions with a total of $(d+2)(d-1)/2$ physical degrees of freedom. However, the field being represented by a symmetric tensor $B_{\mu\nu}$ with $(d+1)(d+2)/2$ free elements, the theory contains $d+2$ degrees of freedom that are spurious. Accordingly, $2(d+2)$ extra constraint equations are required, so as to fix both the values of those fields and their momenta. \\
The most general action containing only  terms of scaling $d+1$ that are quadratic in the rank-2 tensor field $B_{\mu\nu}$ can be written as
\begin{equation}
\begin{split}
    \mathcal{L} = &- \left|D_{\rho} B_{\mu\nu} \right|^2 + 2\left| D_{\mu} B^{\mu\nu}\right|^2 - \left[ D_{\mu} B^{*\mu\nu} D_{\nu} B + h.c. \right] \\
    & + \left| D_{\mu} B \right|^2- m^2 \left( \left|B_{\mu\nu}\right|^2 - \left|B\right|^2 \right) + c_1 R_{\mu\nu\rho\lambda} B^{*\mu\rho}B^{\nu\lambda} \\
    & + c_2 R_{\mu\nu} B^{*\mu \lambda} {B^{\nu}}_\lambda + c_3 \left[ e^{i\theta} R_{\mu\nu} B^{*\mu\nu} B + h.c. \right] \\
    &+ c_4 R \left|B_{\mu\nu}\right|^2 + c_5 R \left|B\right|^2 + i c_6 q F_{\mu\nu} B^{*\mu \lambda} {B^{\nu}}_\lambda.
\end{split}
\label{eq:Lagrangian_BHRY_most_general}
\end{equation}
where the coefficients $c_i$, $i = 1,\dots, 6$ are the Lagrange multipliers that were referred to in Sec.\,\ref{sec:HigherSpinFields} when commenting on the BHRY model.\\
Let $E_{\mu\nu}=0$ denote the EOM for the tensor component $B_{\mu\nu}$. Then, it is possible to divide the equations in the two categories defined before. As it is shown in Ref.\,\cite{Benini:2010pr}, using the usual convention for greek and latin Lorentz indices:
\begin{itemize}
    \item $E_{ij} = 0$ are dynamical equations which determine the second derivative $\ddot{B}_{ij}$.
    \item $E_{\mu t} = 0$ do not contain accelerations. Therefore, they are a set of $d+1$ primary constraints.
\end{itemize}
In order to fix all the degrees of freedom, one applies the conservation of the dynamical equations. This amounts to write $D^{\mu} E_{\mu\nu} =0$. Combined with the trace equation ${E^{\mu}}_{\mu} = 0$, it can be shown that doing so provides an extra set of $d+1$ secondary constraints. That way, only 2 of the required $2(d+2)$ constraints are missing, that can be obtained by iteration. In particular, the penultimate constraint originates from the second divergence $D^{\mu} D^{\nu} E_{\mu\nu} = 0$, combined with the trace equation again. Finally, the last constraint is computed from the time derivative of the first divergence, $D^t D_{\mu} {E^{\mu}}_j = 0$. 

Following these steps, the coefficients $c_1$ to $c_6$ in the most general Lagrangian Eq.\,(\ref{eq:Lagrangian_BHRY_most_general}) can be fixed to give Eq.\,(\ref{eq:Lagrangian:action_BHRY}). This Lagrangian defines a theory for the d-wave holographic superconductor that contains the right number of degrees of freedom. However, as it is discussed once again in Ref.\,\cite{Benini:2010pr}, the theory obtained by this method has several limitations. For instance, the constraint equations only hold if the background is fixed, and so it cannot account for backreactions on the metric.

\subsection{Comparison of the condensates between the na\"{i}ve and physical models}
\label{sec:comparison_naive_physical}
Now that the Abelian-Higgs s-wave, SU(2) Yang-Mills p-wave and CKMWY and BHRY d-wave model have been introduced, everything is set up to resume the comparison between the \emph{na\"{i}ve} and physical models which was started in Sec. \ref{sec:discussion} including Fig.\,\ref{fig:condensates_generalised_model}.

Before comparing the condensates for the \emph{na\"{i}ve} and the mentioned physical models, it is useful to return for a moment to the case of general $\ell$ seeing as it provides some tools to relate the individual models. 
As it has been motivated in Sec. \ref{sec:ansatze_fields} and shown explicitly for $\ell \le 2$, a spin-$\ell$ field can be represented by a tensor of rank $\ell$ accompanied by appropriate constraints.
The general EOM for $\Psi$ with coefficients $E_\ell$ and $F_\ell$ that incorporate the curvature corrections implemented in the physical models has already been given in Eqs.\,(\ref{eq:eom_psi_withEandF}) and (\ref{eq:dynamical_mass_E_F}) in Sec.\ \ref{sec:HigherSpin_Naive_vs_Physical} in terms of $\Psi$.

However, in order to compare with the calculation in Sec.\,\ref{sec:condensate_analytic_calculations}, it is convenient to work in terms of $\psi = \Psi z^\ell$, cf. Eq.\,(\ref{eq:psi_rescaling}). Then the equation reads
\begin{equation}
  \psi''  +  \frac{f'}{f}  \psi' 
     +  \left( \frac{q^2 r_H^2 \phi^2}{z^4 f^2} - \frac{m^2 r_H^2}{z^4 f} + \frac{ \tilde{E}_\ell f'}{zf} + \frac{\tilde{F}_\ell}{z^2} \right) \psi = 0 .
\label{eq:eom_psi_UMF}
\end{equation}
When working in terms of the arbitrary curvature-correction coefficients $\tilde{E}_\ell$ and $\tilde{F}_\ell$,  one finds that, by following the same steps as in Secs.\,\ref{sec:solution_horizon} and\,\ref{sec:solution_asymptotic}, the only differences appear in the coefficients of the expansion near the horizon, which now read

\begin{equation}
   \tilde{\mathcal{A}}_{\ell} = \frac{\eta}{h(1)}, 
\label{eq:horizon:TaylorCoeff_A_general}
\end{equation}
\begin{equation}
    \tilde{\mathcal{B}}_{\ell} = - \left(\frac{\tilde{m}^2}{h(1)} + \tilde{E}_{\ell} \right),
    \label{eq:horizon:TaylorCoeff_B_general}
\end{equation}
\begin{equation}
\begin{split}
   \tilde{\mathcal{C}}_{\ell,1} = \frac{1}{4} &\left[ \tilde{F}_\ell + \tilde{E}_\ell \left(1  + \tilde{E}_\ell\right) + (4 + 2\tilde{E}_\ell)\frac{ \tilde{m}^2}{h(1)}\right.\\
    &\left.+ \frac{\tilde{m}^4}{h^2(1)} + 2h'(1)\frac{\tilde{m}^2}{h^2(1)}   \right],
\end{split}
\label{eq:horizon:TaylorCoeff_C1_general}
\end{equation}
\begin{equation}
    \tilde{\mathcal{C}}_{\ell,2} = -\frac{1}{4h^2(1)},  
\label{eq:horizon:TaylorCoeff_C2_general}
\end{equation}
\begin{equation}
   \tilde\Delta_\ell = \frac{3}{2} + \sqrt{ \left( \frac{3}{2} \right)^2 + \tilde{m}^2 + 2 \tilde{E}_\ell - \tilde{F}_\ell}.
\label{eq:delta_ell_general}
\end{equation}
For the \emph{na\"{i}ve} model, the coefficients are simply given by $\tilde{E}_\ell = 0$ and $\tilde{F}_\ell = -\ell$. Plugging those values in allows to recover the expressions found in Sec.\,\ref{sec:solution_horizon}. 
It is crucial how these coefficients enter into Eqs.\,(\ref{eq:matching:CriticalTemperature})-(\ref{eq:matching:CondensateCoefficient}): the critical temperature and the prefactor of the condensate are modified but the scaling with $\sqrt{1-T/T_c}$ for $T\rightarrow T_c$, that is distinctive of a superconducting condensate, remains unaffected.\\
As it has been discussed in Sec.\,\ref{sec:HigherSpin_Naive_vs_Physical}, the action and the EOM's differ between the \emph{na\"{i}ve} and physical models for p- and d-wave superconductors. Meanwhile, the s-wave models are in agreement with each other. Now, the expressions above allow to compare these differences both at the level of the EOMs and to eventually quantify them.
Starting from Eq.\,(\ref{eq:matching:CondensateFuncTemperature}) and using the updated parameters, Eqs.\, (\ref{eq:horizon:TaylorCoeff_A_general}) - (\ref{eq:delta_ell_general}),
the modifications for the physical models can be regarded as $\tilde{E}_\ell$ and $\tilde{F}_\ell$ terms (cf.\ Eq.\,(\ref{eq:eom_psi_UMF})). Those terms enter $\Delta_\ell$ at the same level as the mass parameter $\tilde{m}^2$. Accordingly, and as it is apparent from the discussion in previous sections, near the AdS-asymptotic boundary they can be absorbed into a dynamical mass 
\begin{equation}
    m_{dyn}^2 = \tilde{m}^2 + 2 \tilde{E}_\ell - \tilde{F}_\ell.
    \label{eq:final_comparison:dynamical_mass}
\end{equation}
Then, the anomalous dimension reads
\begin{equation}
    \Delta_\ell = \frac{3}{2} + \sqrt{\frac{9}{4} + m_{dyn}^2}.
\label{eq:comparison:dynamical_mass_E_F}
\end{equation}
Eq.\,(\ref{eq:final_comparison:dynamical_mass}) implies that, after an appropriate case-wise choice of the value of $\tilde{m}^2$, the values of $m_{dyn}^2$ and thus $\Delta_\ell$ for the \emph{na\"{i}ve} and physical models can easily be matched. 
So, in order to better compare the different models, the asymptotic dynamical mass Eq.\,(\ref{eq:final_comparison:dynamical_mass}) can be chosen at the BF bound, $m_{dyn}^2 = - \frac{9}{4}$, resulting in $\Delta_\ell = \frac{3}{2}$. The corresponding values of $\tilde{m}^2$ together with the parameters $\tilde{E}_\ell$ and $\tilde{F}_\ell$ for each model are presented in Table\, \ref{tab:parameters_comparison_plot}.

\begin{table}[h!]
    \centering
    \begin{tabular}{l|r|r|r}
         Model & $\tilde{E}_\ell$ & $\tilde{F}_\ell$ & $\tilde{m}^2$  \\
    \hline \hline
        \emph{na\"{i}ve} s-wave & 0 & 0 & $-9/4$ \\
    \hline
        \emph{na\"{i}ve} p-wave & 0 & -1 & $-13/4$\\
    \hline
        \emph{na\"{i}ve} d-wave & 0 & -2 & $- 17/4$ \\
    \hline
        \emph{na\"{i}ve} f-wave & 0 & -3 & $- 21/4$ \\
    \hline
        \emph{na\"{i}ve} g-wave & 0 & -4 & $- 25/4$ \\    
    \hline
        Abelian Higgs s-wave & 0 & 0 & $-9/4$ \\
    \hline
        SU(2) YM p-wave & -1 & 0 & $- 1/4$ \\
    \hline
        CKMWY d-wave model & 0 & -1 & $-17/4$ \\
    \hline
        BHRY d-wave & 0 & 0 & $- 13/4$ \\
   \hline
    \end{tabular}
    \caption{Parameter choices for $\tilde{m}^2$ at the BF bound}
    \label{tab:parameters_comparison_plot}
\end{table}

Then, looking back at Fig.\,\ref{fig:condensates_generalised_model}, which shows the condensates as a function of the temperature for these values of $\tilde{m}^2$, the common features as well as the differences are clearly visible. Notice that
since the anomalous operator dimension $\Delta_\ell$ agrees for all models, using the normalisation in the figure all curves exhibit the same scaling with the temperature. The only differences originate from the coefficient $W_\ell$ of the condensate defined by Eq.\,(\ref{eq:matching:CondensateCoefficient}), that results in an overall multiplicative factor for the curves.

Naturally, due to the scalar nature of the matter field and the simplicity of its description, the \emph{na\"{i}ve} s-wave model agrees with the Abelian-Higgs s-wave model and so do the respective condensates. Meanwhile, the two curves for \emph{na\"{i}ve} and SU(2) Yang-Mills p-wave superconductors coincide because, by chance, the coefficient $W_\ell$ is the same for both models and the changes in $\mathcal{B}_\ell$ due to the different choice of $\tilde{m}^2$ are cancelled equally by $\tilde{E}_\ell$ in Eq.\,(\ref{eq:horizon:TaylorCoeff_B_general}). 

This is no longer the case for the d-wave superconductor, since $\mathcal{B}_\ell$ and thus $W_\ell$ agree between the \emph{na\"{i}ve} and the CKMWY models (which coincide at the level of the action), but they both differ from the BHRY model. Even so, as it has already been stated, the differences appear only in the overall factor of the condensate. By taking the ratio of the coefficients $W_\ell$ for the \emph{na\"{i}ve} and BHRY p-wave model, 
\begin{equation}
    \frac{W_p^\text{BHRY}}{W_p^\text{naîve}} \approx 1.258,
\end{equation}
one finds that, from a phenomenological point of view, the models differ by the order of a 25\% in their prediction of the magnitude of the superconducting condensate.

\section{Conclusions}

In this paper we presented a generalisation of the holographic approach towards superconductors.\\
First, a \emph{na\"{i}ve} model that describes the action of a matter 
field with spin of arbitrary value $\ell$, coupled to an electromagnetic field, has been introduced in the bulk. 
Afterwards, the widely-used AdS Schwarzschild Black Hole metric has been generalised to sets of metrics with a first-order zero at the horizon.\\
Eventually, equations of motion for the relevant degrees of 
freedom of the theory have been solved by means of the 
semi-analytical matching approach presented in Ref.~\cite{Gregory:2009fj}.\\
One great virtue of the \emph{na\"{i}ve} model consists in the fact that it can reproduce the expected scaling of the condensate with the temperature, $\mathcal{O} \sim \sqrt{1-T/T_c}$, independently of $\ell$. Notice that this was not at all straightforward, as 
the \emph{na\"{i}ve} model is not a self-consistent field theory
in $d+1$-dimensional curved spacetime for any value of $\ell$: 
although the model agrees with the Fronsdal equations in Minkowski spacetime, it does neither account for correction terms in curved spacetime nor describe the correct amount of physical degrees of freedom. It hence needs to be endowed with suitable constraint equations, in order to be promoted to a consistent \emph{physical} field theory. \\
In spite of this, the results for the condensate and the critical temperature retain the correct dependence on the free parameters of the theory and allow for an analytical study for any value of $\ell$
considered in the literature so far.\\
In particular, the mentioned model reproduces the Abelian-Higgs model for s-wave superconductors and the CKMWY model for d-wave superconductors studied in Ref.\,\cite{Hartnoll:2008kx} and  Ref.\,\cite{Chen:2010mk}, respectively. 
On the other hand, the equations of motion derived in the
\emph{na\"{i}ve} model differ slightly from the SU(2) Yang-Mills model for p-wave superconductors of Ref.\,\cite{Gubser:2008wv} and the BHRY model for d-wave superconductors of Ref.\,\cite{Benini:2010pr}. 
The differences can be regarded as a change in the dynamical mass and only affect the critical temperature's dependence on the extra-dimensional parameters, whereas the scaling with the charge density $\rho$ is universal and leads to the same  
superconductor phenomenology. Moreover, a direct comparison of the numerical results shows that the coefficient of the condensate computed with the \emph{na\"{i}ve} d-wave model differs by approximately 25\% from the \emph{physical} BHRY d-wave model.\\
In order to address the important point of the difference
in the equations of motion for the relevant
degrees of freedom that arises when comparing the \emph{na\"{i}ve} model and the consistent field theory for a given spin $\ell$,
a technique to compute the constraint equations has been summarised. For the d-wave superconductors the constraint equations were calculated explicitly to shed some light on how to obtain the \emph{physical} equations of motions from the action in Eq.\,(\ref{eq:general_matter_Lagrangian}). We found that
in all cases studied in the literature the equations of motion
derived using the \emph{na\"{i}ve} model 
single out the essential features a holographic model for superconductivity must encapsulate.\\
Our results for s-, p- and d-wave holographic superconductors
encourage us to speculate about the behaviour of possible 
higher-spin superconductors.
In recent works~\cite{won2000possible}, the existence of f-wave superconductivity was conjectured on systems like Sr$_2$RuO$_4$. More recently~\cite{wu2021nature}, f-wave superconductivity was observed in AV$_3$Sb$_5$, where A=K, Rb or Cs. Using our general description in the \emph{na\"{i}ve} model for any $\ell$, we have been able to describe new types of possible superconductivity behaviour, including the f-wave case, which could help describing these new systems.   

\section*{Acknowledgements}
VS would like to thank Daniel Errandonea for pointing out relevant works in f-wave superconductivity, and F\"eanor Reuben Mitchell Ares for discussions on the analytical matching procedure. This work was supported by the grant PROMETEO/2019/08 and the Spanish FPA2017-85985-P. FE is supported by the Generalitat Valenciana with the grant GRISOLIAP/2020/145. VS  acknowledges  support from the UK Science and Technology Facilities Council ST/L000504/1. This project  has  received funding support  from  the  European Union’s Horizon 2020 research and innovation programme under the Marie Sklodowska-Curie grant agreement 860881-HIDDeN. \\
A preprint has previously been published on arXiv \cite{thiswork}.

\appendix

\section{Higher order singularities at the horizon}
\label{sec:appendix_generalised_metric}

In Sec.\,\ref{sec:generalised_metric}, the background metric has been generalised to the form introduced by Eq.\,(\ref{eq:generalized_f_metric}) in an effort to fully generalise the description of holographic superconductors. However, the function $f(z) \propto (1-z)$ in  Eq.\,(\ref{eq:generalized_f_metric}) defines the singularity at the horizon $z = 1$ to be of first order. One might then ask whether functions of the more general form
\begin{equation}
    f(z) = \frac{r_H^2}{L^2} h(z) (1-z)^{1+\alpha}
    \label{sec:appendix_alpha:f_alpha}
\end{equation}
could also give rise to the desired superconducting behaviour. In this appendix, it will be briefly argued that the case $\alpha = 0$ provides in fact the simplest scenario for the phenomenological description of superconductors. 

To look at this issue, one may reproduce the steps in Sec.\ \ref{sec:condensate_analytic_calculations} that lead to the analytical calculation of the superconducting condensate. When examining the regularity conditions near the horizon,  Eq.\,(\ref{eq:horizon:PsiBoundaryCondition}) is modified into
\begin{equation}
    \psi'(1) \equiv \lim_{z \to 1}\psi'(z) = \lim_{z \to 1} - \frac{\tilde{m}^2 (1-z)^{-\alpha}}{p z^4 h(z)} \psi(z),
\label{eq:horizon_psi_boundary_condition_p}    
\end{equation}
where $\psi$ is given by Eq.\,(\ref{eq:psi_rescaling}) and $h(z)$ is regular at $z = 1$.

Starting with $-1 <\alpha < 0$, this implies that there are two cases: 
\begin{itemize}
    \item $\psi'(1) = 0$: then, in order to satisfy the EOM for $\psi$ at the horizon, that is analogous to Eq.\,(\ref{eq:eom_psi}) in the case $\alpha = 0$, $\psi(1) = 0$ must be satisfied. As a consequence, one also finds $\psi''(1) = 0$ using the Taylor expansion near the horizon. 
    This means that the matter field $\psi$ disappears up to second order at the horizon and the condensate vanishes.
 
   \item $\psi(1)\rightarrow \infty$: let $\psi$ be given by a power-law solution of the form $\psi(z) \sim (1-z)^{-\beta}$ near the horizon, then it is $\psi'(z) \sim (1-z)^{-(\beta+1)}$. Therefore, $\psi'(z)$ scales with a larger power of $(1-z)$, contradicting Eq.\,(\ref{eq:horizon_psi_boundary_condition_p}). 
\end{itemize}
In both cases, $\psi$ does not contain a quadratic term ${(1-z)^2}$ which, as can be inferred from the results in Sec.\,\ref{sec:condensate_analytic_calculations}, is one crucial ingredients
for the emergence of a condensate. Consequently, there is no condensate in this case.

For $\alpha >0$ there are similarly two options:
\begin{itemize}
    \item $\psi(1) = 0$: if that is the case, the EOM  for $\psi$ similar to Eq.\,(\ref{eq:eom_psi}) enforces $\psi'(1) = 0$ and, after the Taylor expansion Eq.\,(\ref{eq:horizon:PsiTaylorExpansion}), also $\psi''(1) = 0$. Again, $\psi$ vanishes again up to second order at the horizon.
    \item $\psi'(1)\rightarrow \infty$: if $\psi$ had a power-law form ${\psi(z) = (1-z)^\beta}$, then $\psi'(z) \propto (1-z)^{\beta-1}$. This implies $\alpha = 1$ by Eq.\,(\ref{eq:horizon_psi_boundary_condition_p}). I.e. for $p =2$ solutions could exist, but no analytical solution could be found in this case. 
\end{itemize}
Therefore, no condensate emerges when $\alpha>0$, either, for the family of solutions under consideration. To summarise then, the condensate vanishes for $\alpha \neq 0$, at least when considering the family of solutions obtained from power series by enforcing regularity conditions on the fields.

\section{Dependence of the analytical results on the matching point}\label{sec:depzm}

\begin{figure}[h!]
    \centering
 \includegraphics[width = 0.5 \textwidth]{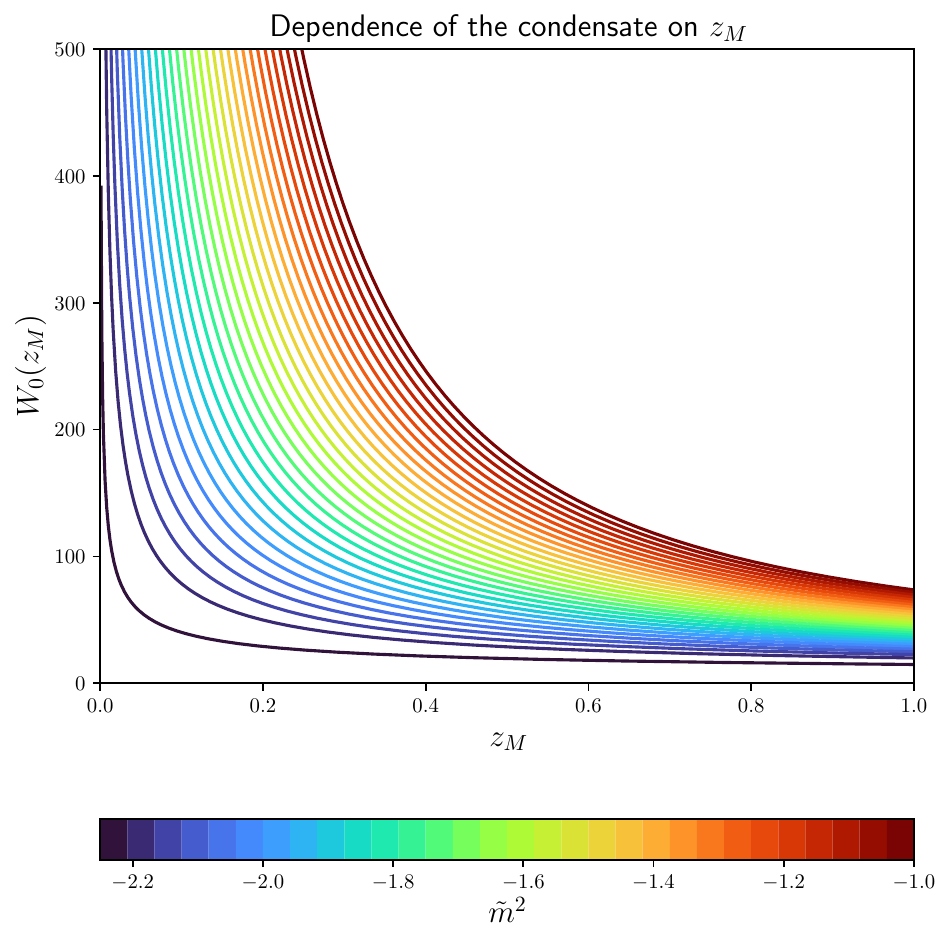}
    \caption{Dependence of the coefficient $W_\ell(z_M)$ of the condensate on the matching point $z_M$ for an s-wave superconductor ($\ell = 0$). It is remarkable that for larger $\tilde{m}^2$ (thus increased $\Delta_\ell$), the choice of the matching point becomes irrelevant, provided it is far enough from the pole at $z = 0$. Note that the BF-bound for $\ell=0$ is situated at $\tilde{m}^2 = -2.25$.}
    \label{fig:z_Match_dependence}
\end{figure}

In Sec.\,\ref{sec:condensate_analytic_calculations}, an analytical method was presented that allows to disentangle the dynamics of holographic superconductors. Said method consists in the construction of asymptotic solutions of the EOMs of a theory both near the event horizon and the AdS-asymptotic region, which are then glued together at an intermediate point. In this appendix, a technical aspect of this calculation is briefly clarified: the importance of the choice of the matching point $z = z_M$ to the coefficient of the condensate.

\begin{figure*}[ht]
    \centering
    \includegraphics[width=\textwidth]{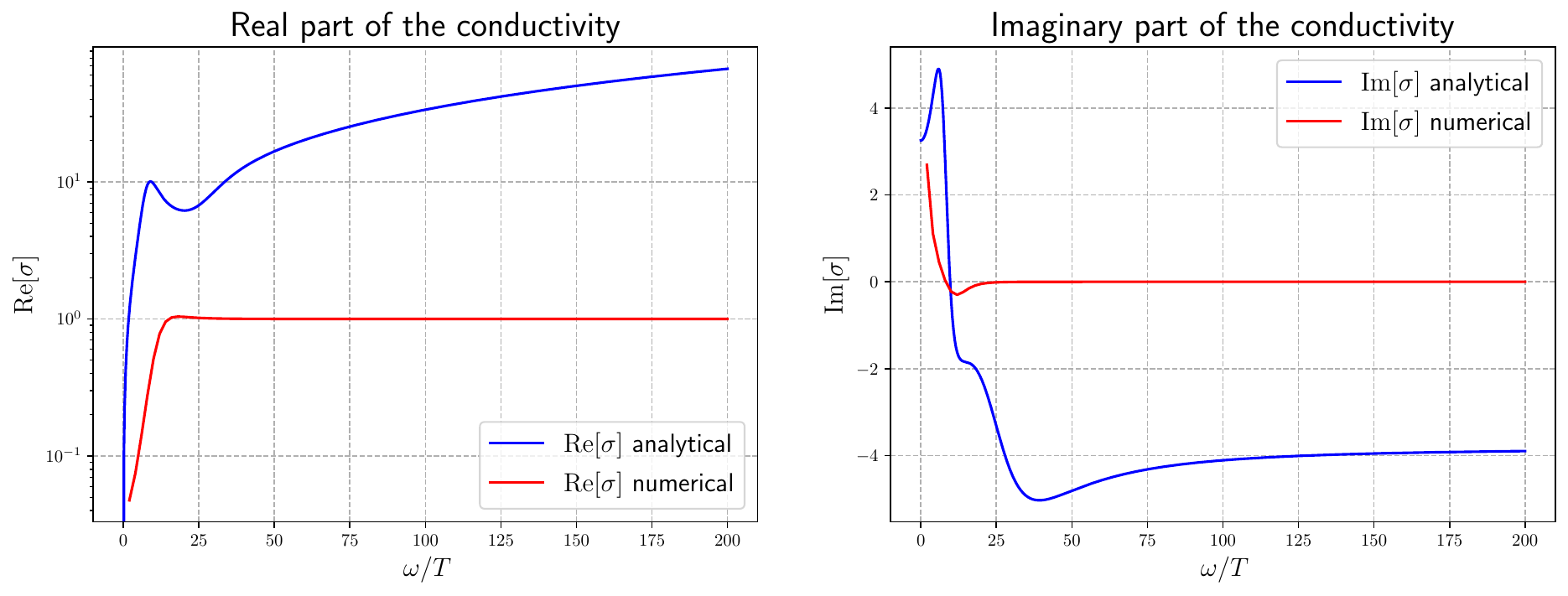}
    \caption{Comparison of the real and imaginary part of the conductivity for the analytical and numerical approach. Note that the real part of the conductivity is plotted logarithmically while the plot of the imaginary part employs a linear scale. Notice also how in the low frequency region, the analytical $\sigma$ reproduces the features of the numerical one to some extent. However, its behaviour is completely faulty for $\omega>\omega_g$.}
    \label{fig:conductivites}
\end{figure*} 

Keeping an explicit dependence on $z_M$, the condensate reads 
\begin{equation}
    \langle \mathcal{O} \rangle = W_\ell(z_M) T_c T^{l+\Delta -1} \sqrt{1+\frac{T}{T_c}}\sqrt{1-\frac{T}{T_c}},
\end{equation}
where the $z_M$ dependent coefficient is given by 
\begin{equation}
    W_\ell(z_M) = \left(\frac{4\pi}{3}\right)^{\ell + \Delta_\ell}\frac{\sqrt{2}z_M^{1-\Delta_\ell}[2+\mathcal{B}_\ell(z_M-1)]}{\sqrt{\mathcal{A}_\ell}[\Delta_\ell-z_M(\Delta_\ell-2)]}
\label{eq:general_matching_Wl}
\end{equation}
and $\mathcal{A}_\ell$ and $\mathcal{B}_\ell$ are the coefficients defined by Eqs.\,(\ref{eq:horizon:TaylorCoeff_A}) and\,(\ref{eq:horizon:TaylorCoeff_B}), respectively. This expression coincides with Eq.\,(\ref{eq:matching:CondensateCoefficient}) when $z_M = 1/2$. 

The only problematic point within the range of definition of $z_M$ (recall that $z\in (0, 1]$)
is the pole which arises at $z_M = 0$ whenever $\Delta_\ell>0$. 
Notably, as Eq.\,(\ref{eq:general_matching_Wl}) shows, the pole scales as $z_M^{1-\Delta_\ell}$.
Therefore, the smaller the scaling dimension of the condensate $\Delta_\ell$ is, the steeper and narrower is the growth of $W_\ell(z_M)$ towards the pole at $z_M = 0$, making the overall dependence with $z_M$ milder. This behaviour, which can be spotted in Fig.\,\ref{fig:z_Match_dependence}, justifies the choice of the value of $\tilde{m}^2$ that saturates the BF-bound for an optimal performance of the semi-analytical calculation method that has been used in this paper.

\section{The electric conductivity}
\label{sec:appendix_conductivity}

Two different quantities typically characterise holographic superconductors. One is the condensate, whose emergence has been readily checked in this paper using the analytical matching method presented in Sec.\,\ref{sec:condensate_analytic_calculations}. Additionally, the optical conductivity is worth being considered, since it generically exhibit a recognisable gap at some frequency $\omega_p$ in superconducting systems. With this as its outstanding feature, the conductivity can be used to classify systems as superconductors.

However, little attention has been brought to the conductivity throughout this paper. The reason for the lack of focus on that quantity is that, as it will be shown in this last appendix, the calculation method that has been made use in this paper breaks down when an analytical computation of the conductivity is attempted. 

Nevertheless, we find it illustrative to take a look at the conductivity to show that the presented matching approach could provide a suitable starting point to derive semi-analytical calculation methods for the conductivity.
The modification of the analytical method to account for the conductivity is left for a future work.

In order to understand why this breaking happens, the results emerging from the analytical method will be put in comparison with the numerical outcome for the simple case of the s-wave superconductor, $\ell = 0$. 

So as to obtain the conductivity, one starts by considering the fluctuations of the EM gauge field in a direction transversal to the AdS-radial dimension $r$. Since the s-wave superconductor is rotationally invariant in its spatial dimensions, it is sufficient to consider fluctuations in the x-direction only:
\begin{equation}
\delta A = A_x(r) e^{-i\omega t} dx.    
\end{equation}
Then, the EOM for this fluctuation in a generic BH background reads
\begin{equation}
    A_x^{''} + \frac{f^{'}}{f}A_x^{'} + \left( \frac{\omega^2}{f^2} - \frac{2\Psi^2}{f} \right) A_x = 0.
\label{eq:EOM_Ax}
\end{equation}
As described in Refs.\,\cite{Hartnoll:2008kx, Hartnoll:2008vx}, numerical solutions to this equation can be straight-forwardly obtained once the EOMs for $\psi$ and $\phi$ have been solved. The resulting numerical conductivity $\sigma$ then displays the expected behaviour, as it can be observed in Fig.\,\ref{fig:conductivites}, where its real and imaginary parts are represented in red. In particular, its real part is negligible for smaller frequencies $\omega<\omega_p$ and becomes asymptotically unity as the frequency surpasses $\omega_p$, shaping the aforementioned gap around that frequency. 

Then, motivated by its successful description of the condensates (c.f. Sec.\,\ref{sec:condensate_analytic_calculations}), one may attempt to apply the same analytical method to the calculation of the conductivity. For that purpose, the gauge field fluctuations near the horizon can be approximated by
\begin{equation}
\begin{split}
    A_x^H = &\left[3r_H (1-z)\right]^{- \frac{i\omega}{3 r_h}} \times \\
    &\left[1+ A_x^a (1-z) + A_x^b (1-z^2) + A_x^c (1-z)^3\right],
\end{split}
\label{eq:Ax_hor}
\end{equation}
since near the horizon, $A_x \sim f^\alpha$ with $\alpha = \pm\frac{i\omega}{3 r_h}$ while the metric asymptotically becomes $f_{BH} \longrightarrow 3r_H (1-z)$. \\
Afterwards, the coefficients $A_x^a$, $A_x^b$ and $A_x^c$ are fixed by expanding and solving the EOM 
Eq.\,(\ref{eq:EOM_Ax}) up to ${\mathcal{O}(1-z)^2}$.

On the other hand, in the asymptotic region the fluctuations can be written as 
\begin{equation}
    A_x^{AdS} = A_x^0 + \frac{A_x^1}{r} + ...
\label{eq:Ax_asymp}
\end{equation}
As done in the case of the condensate, the
expansion near the horizon, Eq.\,(\ref{eq:Ax_hor}), is matched to the asymptotic expansion near the AdS boundary, Eq.\,(\ref{eq:Ax_asymp}), by imposing Dirichlet and von Neumann boundary condition at the matching point $z_M$, i.e.
\begin{equation}
    \begin{split}
        A_x^H(z_M) &= A_x^{AdS}(z_M) \\
        A'^H_x(z_M) &= A'^{AdS}_x (z_M).
    \end{split}
\end{equation}
These two equations determine the coefficients $A_x^0$ and $A_x^1$. According to e.g.\ Ref.\,\cite{Hartnoll:2008kx}, the conductivity can be defined in terms of those coefficients as

\begin{equation}
    \sigma = \frac{1}{i\omega} \frac{A_x^1}{A_x^0}.
\label{eq:conductivity}
\end{equation}

However, when the analytical coefficients are plugged in this definition, the outcome is very far from resembling the reliable numerical solution described before. This is also to be seen in Fig.\ \ref{fig:conductivites}, where the real and imaginary part of the analytically calculated conductivity are put in comparison with the numerical outcome.
In fact, one can see that while some gap-like structure develops for $\omega$\,$<$\,$\omega_p$, the analytical conductivity does not retain its proper normalisation and grows indefinitely as $\omega\longrightarrow\infty$. This is a non-physical behaviour in what respects the description of superconductors that indicates the breaking of this method of calculation. 

To understand the reason for this failure, one may look at the expansion in Eq.\,(\ref{eq:Ax_hor}), whose prefactor forecasts an oscillating behaviour of the solution near the horizon. This oscillating solution is then matched to an a-priori non-oscillating expansion near the asymptotic boundary, Eq.\,(\ref{eq:Ax_asymp}). Therefore, a high dependence on the matching point $z_M$ is expected that spoils any hope of finding a proper expression for the conductivity using this method. Nevertheless, more refined matching procedures, beyond the one employed in this paper, may lead to a successful analytical treatment of the conductivity properties of SCs. We leave this study for a future work.

\bibliography{HoloBibliography.bib}

\end{document}